\documentclass[conference]{IEEEtran}
\usepackage{amsmath}
\usepackage{amssymb}
\usepackage{amsthm}
\usepackage{booktabs}
\usepackage[compatibility=false]{caption}
\usepackage{color}
\usepackage{enumerate}
\usepackage[T1]{fontenc}
\usepackage{graphicx}
\usepackage{ifthen}
\usepackage{mathtools}
\usepackage{multicol}
	\usepackage{pdflscape}
	\usepackage{rotating}
\usepackage{xspace}
\usepackage[table]{xcolor}
\usepackage{pgf}
\usepackage{tikz}
\usetikzlibrary{arrows,automata,calc,matrix,positioning,shadows}
\usepackage[underline=false]{pgf-umlsd}

\ifCLASSOPTIONcompsoc
  \usepackage[nocompress]{cite}
\else
  \usepackage{cite}
\fi

\newcommand{\eg}{\textrm{e.g.,}\@\xspace}

\newcommand{\ie}{\textrm{i.e.,}\@\xspace}

\newcommand{\etc}{\textrm{etc.}\@\xspace}

\newtheorem{definition}{Definition}

\newtheorem{proposition}{Proposition}

\newtheorem{example}{Example}[section]

\usepackage{datetime}

\newcommand{\lnotation}[4]{
	\def\third:{#3} 
	\def\possiblyone:{} 
	\def\possiblytwo:{~}
	\def\possiblythree:{ }
	\def\divide{\;#1\hspace*{-0pt}( #2\; \mid: \; #4 \, )}
	\def\nodivide{\;#1\hspace*{-0pt}( #2\;\mid\; #3\;:\;#4 \, )}
	\ifx\third\possiblyone\divide
		\else\ifx\third\possiblytwo\divide
		\else \ifx\third\possiblythree\divide
		\else \nodivide\fi\fi\fi}

\newcommand{\biglnotation}[4]{
	\def\third:{#3} 
	\def\possiblyone:{} 
	\def\possiblytwo:{~}
	\def\possiblythree:{ }
	\def\divide{\;#1\hspace*{-0pt}\big( #2\; \mid: \; #4 \, \big)}
	\def\nodivide{\;#1\hspace*{-0pt}\big( #2\;\mid\; #3\;:\;#4 \, \big)}
	\ifx\third\possiblyone\divide
		\else\ifx\third\possiblytwo\divide
		\else \ifx\third\possiblythree\divide
		\else \nodivide\fi\fi\fi}

\newcommand{\bigglnotation}[4]{
	\def\third:{#3} 
	\def\possiblyone:{} 
	\def\possiblytwo:{~}
	\def\possiblythree:{ }
	\def\divide{\;#1\hspace*{-0pt}\bigg( #2\; \mid: \; #4 \, \bigg)}
	\def\nodivide{\;#1\hspace*{-0pt}\bigg( #2\;\mid\; #3\;:\;#4 \, \bigg)}
	\ifx\third\possiblyone\divide
		\else\ifx\third\possiblytwo\divide
		\else \ifx\third\possiblythree\divide
		\else \nodivide\fi\fi\fi}
 
\newcommand{\wplnotation}[4]{
	\def\third:{#3} 
	\def\possiblyone:{} 
	\def\possiblytwo:{~}
	\def\possiblythree:{ }
	\def\divide{\;#1\hspace*{-0pt}\bigg( #2\; \mid: \; #4 \, }
	\def\nodivide{\;#1\hspace*{-0pt}\bigg( #2\;\mid\; #3\;:\;#4 \, }
	\ifx\third\possiblyone\divide
		\else\ifx\third\possiblytwo\divide
		\else \ifx\third\possiblythree\divide
		\else \nodivide\fi\fi\fi}

\newcommand{\grieslnotation}[4]{
	\def\third:{#3} 
	\def\possiblyone:{} 
	\def\possiblytwo:{~}
	\def\possiblythree:{ }
	\def\divide{(#1 #2\; \mid : \; #4 \, )}
	\def\nodivide{(#1 #2\;\mid\; #3\;:\;#4 \, )}
	\ifx\third\possiblyone\divide
		\else\ifx\third\possiblytwo\divide
		\else \ifx\third\possiblythree\divide
		\else \nodivide\fi\fi\fi}

\newcommand{\griesset}[3]{
	\def\second:{#2} 
	\def\possiblyone:{} 
	\def\possiblytwo:{~}
	\def\possiblythree:{ }
	\def\divide{\{#1\; \mid : \; #3 \, \}}
	\def\nodivide{\{#1\;\mid\; #2\;:\;#3 \, \}}
	\ifx\second\possiblyone\domvide
		\else \ifx\second\possiblytwo\divide
		\else \ifx\second\possiblythree\divide
		\else \nodivide\fi\fi\fi}

\newcommand{\set}[1]{\{#1\}}

\newcommand{\sets}[2]{\{#1\; \mid \; #2\}}

\newcommand{\bigsets}[2]{\big\{#1\; \mid \; #2\big\}}

\newcommand{\A}{{\mathcal{A}}}

\newcommand{\monoid}[3]{\big(#1, #2, #3 \big)}

\newcommand{\semiring}[5]{\big(#1, #2, #3, #4, #5\big)}

\newcommand{\Lsemimodule}[3]{\big(_{#1}#2, #3\big)}
\newcommand{\Rsemimodule}[3]{\big(#2_{#1}, #3\big)}

\newcommand{\bigP}[1]{\big( #1 \big)}

\newcommand{\bigB}[1]{\big[ #1 \big]}

\newcommand{\bigA}[1]{\big\langle #1 \big\rangle}

\newlength{\interligne}

\newcommand{\Beginproof}{\dimen123=\linewidth \dimen124=\linewidth
	\advance\dimen123 by -15mm \advance\dimen124 by -5mm
	\advance\dimen123 by -\parindent \advance\dimen124 by -\parindent
	\setlength{\interligne}{\baselineskip}
	\setlength{\baselineskip}{1.2\baselineskip}
    	\begin{tabbing}
    		\hspace*{\parindent}\= \hspace*{5mm}\= \kill \+ \kill}
			\newcommand {\Endproof}
		{\end{tabbing}
    \setlength{\baselineskip}{\interligne}}

\newcommand{\com}[1]{\> \hspace*{5mm}
	$\langle$~\parbox[t]{\dimen123}{ #1 $\rangle$}\\}

\newcommand{\pred}[1]{\>\parbox[t]{\dimen124}{#1}\\}

\newcommand{\hsep}{\quad\&\quad}

\newcommand{\Beginproofitem}{\dimen123=\linewidth \dimen124=\linewidth
	\advance\dimen123 by -20mm \advance\dimen124 by -5mm
	\advance\dimen124 by -\parindent
    	\begin{tabbing}
    		\hspace*{5mm}\= \kill}
			\newcommand {\Endproofitem}
	{\end{tabbing}}

\newcommand {\Begingtabin}{
	\begin{tabbing}
    	\hspace*{\parindent}\= \kill \+ \kill}
		\newcommand {\Endtabin}
	{\end{tabbing}}

\newcommand{\Begingspec}{
	\begin{tabbing}
    	\hspace*{\parindent}\= \hspace*{5mm}\=\hspace*{5mm}\=\hspace*{5mm}\=
			\hspace*{5mm}\=\hspace*{5mm} \kill \+ \kill}
		\newcommand {\Endspec}
	{\end{tabbing}}

\newcommand {\Beginspecitem}{
	\begin{tabbing}
    	\hspace*{5mm}\=\hspace*{5mm}\=\hspace*{5mm}\=
    	\hspace*{5mm}\=\hspace*{5mm} \kill}
		\newcommand {\Endspecitem}
	{\end{tabbing}}

\newcommand{\nln}{@{}l@{}}

\newcommand{\ShowEqSourceStructure}{1}

\newcommand{\eqfrom}[1]{
	\ifthenelse{\ShowEqSourceStructure=1}
	{ \quad {\textcolor{red}{\mbox{#1}}}} {\quad} }

\newcommand{\true}{\textsf{true}}
\newcommand{\false}{\textsf{false}}

\newcommand{\Or}{\mathrel{\vee}}
\newcommand{\Ors}{\;\Or\;}
\newcommand{\AAnd}{\mathrel{\wedge}}
\newcommand{\nAnd}{\;\AAnd\;}

\newcommand{\mImp}{\;\Longrightarrow\;}  
\newcommand{\mImpl}{\;\Longleftarrow\;}  
\newcommand{\mIff}{\;\Longleftrightarrow\;}

\newcommand{\STbot}{\emptyset}

\newcommand{\STleq}{\subseteq}

\newcommand{\STjoin}{\; \cup \;}
\newcommand{\STmeet}{\; \cap \;}
\newcommand{\STdiff}{\backslash}

\newcommand{\internalconverse}[1]
	{#1^{\mkern-1mu{}{\raise0.1ex\hbox{\tiny$\smallsmile\,$}}}\kern-0.1em{}}

\newcommand{\RAcomp}{\mathop{\kern-.5pt\raise.3ex\hbox{\footnotesize\rm;}}}

\newcommand{\guardsymb}{\lceil\!\:\!\!\rfloor}
\newcommand{\guardif}[2]{\ifs \quad #1 \longrightarrow #2}
\newcommand{\guardb}[2]{\guardsymb\; \quad #1 \longrightarrow #2}
\newcommand{\guarde}{\fis}

\newcommand{\abs}[1]{|#1|}

\newcommand{\deq}{\spaces{\stackrel{\mathrm{def}}{=}}}

\newcommand{\keyw}[1]{{\textsf{#1}}}

\newcommand{\rspace}[1]{#1\ }

\newcommand{\spaces}[1]{\,#1\,}

\newcommand{\fis}{\keyw{fi}}

\newcommand{\ifs}{\rspace{\keyw{if}}}

\newcommand{\receive}{\rspace{\keyw{receive}}}

\newcommand{\nulls}{\keyw{null}}

\newcommand{\var}[1]{\texttt{\small{#1}}}
\newcommand{\varL}[1]{\texttt{#1}}

\newcommand{\KA}{Kleene algebra\@\xspace}

\newcommand{\CKA}{concurrent Kleene algebra\@\xspace}
\newcommand{\CKAabbrv}{\textup{CKA}\@\xspace}

\newcommand{\CKAset}{K}

\newcommand{\KAstar}[1]{{#1}^*}
\newcommand{\CKApar}{*}
\newcommand{\CKAseq}{\raise.3ex\hbox{\,\rm;\,}}
\newcommand{\CKAiterSeqOp}{\text{\scriptsize \textcircled{\raise.3ex\hbox{\,\rm;\,}}}}
\newcommand{\CKAiterParOp}{\text{\scriptsize \textcircled{\raise-.75ex\hbox{\,*\,}}}}
\newcommand{\CKAiterSeq}[1]{{#1}^\CKAiterSeqOp}
\newcommand{\CKAiterPar}[1]{{#1}^\CKAiterParOp}

\newcommand{\KAstructure}{\bigP{\CKAset, +, \cdot, \KAstar{}, 0, 1}}
\newcommand{\KAsemiring}{\semiring{\CKAset}{+}{\cdot}{0}{1}}

\newcommand{\cka}{{\mathcal K}}

\newcommand{\CKAstructure}{\bigP{\CKAset, +, \CKApar, \CKAseq, \CKAiterPar{}, \CKAiterSeq{}, 0, 1}}
\newcommand{\CKAstructurePar}{\bigP{\CKAset, +, \CKApar, \CKAiterPar{}, 0, 1}}
\newcommand{\CKAstructureSeq}{\bigP{\CKAset, +, \CKAseq, \CKAiterSeq{}, 0, 1}}

\newcommand{\CKAle}{\le_{\cka}}

\newcommand{\STIMset}{S}

\newcommand{\STIMplus}{\oplus}
\newcommand{\STIMdot}{\odot}

\newcommand{\Nstim}{\mathfrak{n}}

\newcommand{\Dstim}{\mathfrak{d}}

\newcommand{\stim}{{\mathcal S}}
\newcommand{\STIMstructure}{\bigP{\STIMset, \STIMplus, \STIMdot, \Dstim, \Nstim}}

\newcommand{\STIMle}{\le_{\stim}}

\newcommand{\STIMevent}[1]{\mathit{#1}}

\newcommand{\rightSemimodule}[1]{right~$#1$-semimodule\@\xspace}
\newcommand{\leftSemimodule}[1]{left~$#1$-semimodule\@\xspace}

\newcommand{\actOp}{\circ}

\newcommand{\lActSig}{\actOp: \STIMset \times \CKAset \to \CKAset}

\newcommand{\lAct}[2]{{#2} \actOp {#1}}

\newcommand{\outOp}{\lambda}			

\newcommand{\lOutSig}{\outOp: \STIMset \times \CKAset \to \STIMset}	
\newcommand{\lOut}[2]{\outOp(#2,#1)}	

\newcommand{\lOutbig}[2]{\outOp\bigP{#2,#1}}

\newcommand{\CCKA}{Communicating Concurrent Kleene Algebra\@\xspace}
\newcommand{\CCKAabbrv}{\textup{C$^2$KA}\@\xspace}

\newcommand{\CCKAstructure}{\bigP{\stim, \cka}}

\newcommand{\ActSemimodule}{\Lsemimodule{\stim}{\CKAset}{+}}
\newcommand{\OutSemimodule}{\Rsemimodule{\cka}{\STIMset}{\STIMplus}}

\newcommand{\levelOne}{stimulus-response specification\@\xspace}
\newcommand{\levelONE}{Stimulus-Response Specification\@\xspace}
\newcommand{\levelTwo}{abstract behaviour specification\@\xspace}
\newcommand{\levelTWO}{Abstract Behaviour Specification\@\xspace}
\newcommand{\levelThree}{concrete behaviour specification\@\xspace}
\newcommand{\levelTHREE}{Concrete Behaviour Specification\@\xspace}

\newcommand{\behave}[1]{\textsc{#1}}

\newcommand{\Agent}[1]{\mathsf{#1}}

\newcommand{\agent}[2]{\Agent{#1} \mapsto \bigA{#2}} 

\newcommand{\STIMbasic}{\STIMset_{a}}
\newcommand{\CKAbasic}{\CKAset_{a}}

\newcommand{\comm}[2]{\mathrel{{\to}_{#1}^{#2}}}

\newcommand{\STIMcommD}[2]{#1 \comm{\stim}{} #2}

\newcommand{\env}{{\mathcal{E}}}
\newcommand{\ENVcommD}[2]{#1 \comm{\env}{} #2}

\newcommand{\depOp}{\mathrm{R}}

\newcommand{\dep}[2]{#2 \,\depOp\, #1}

\newcommand{\toS}{\comm{\stim}{}}
\newcommand{\toE}{\comm{\env}{}}

\newcommand{\typeX}[1]{{\mathcal{T}}_{#1}}
\newcommand{\AgentX}[1]{\Agent{A}_{#1}}
\newcommand{\toX}[1]{\to_{\typeX{#1}}}
\newcommand{\impN}{\AgentX{n} \toX{n} \AgentX{n-1} \toX{n-1} \dots \toX{2} \AgentX{1} \toX{1} \AgentX{0}}
\newcommand{\impX}[1]{p_{#1}^{\typeX{#1}}}
\newcommand{\impS}[1]{p_{#1}^{\stim}}
\newcommand{\impE}[1]{p_{#1}^{\env}}

\newcommand{\infl}[1]{\mathrm{Infl}(#1)}

\newcommand{\defV}[1]{\mathrm{Def}(#1)}
\newcommand{\refV}[1]{\mathrm{Ref}(#1)}

\newcommand{\attackS}[1]{\mathrm{AS}\bigP{#1}}
\newcommand{\attackV}[1]{\mathrm{AV}\bigP{#1}}
\newcommand{\attack}[1]{\mathrm{attack}\bigP{#1}}

\newcommand{\exploit}[1]{\xi\bigP{#1}}

\newcommand{\divRel}{\mathrel{\mathsf{div}}}

\newcommand{\dent}[1]{
	\ifthenelse{\equal{#1}{0}}{\qquad\quad}{}
	\ifthenelse{\equal{#1}{1}}{\qquad\quad}{}
	\ifthenelse{\equal{#1}{2}}{\qquad\qquad\qquad\qquad}{}
	\ifthenelse{\equal{#1}{3}}{\qquad\qquad\qquad\qquad\qquad\qquad}{}
}

\newcommand{\portcaptain}{Port Captain\@\xspace}
\newcommand{\shipmanager}{Ship Manager\@\xspace}
\newcommand{\stevedore}{Stevedore\@\xspace}
\newcommand{\terminalmanager}{Terminal Manager\@\xspace}
\newcommand{\crane}{Crane Manager\@\xspace}
\newcommand{\carrier}{Carrier Coordinator\@\xspace}

\newcommand{\PC}{\Agent{PC}}
\newcommand{\SM}[1]{\Agent{SM}_{#1}}
\newcommand{\SV}[1]{\Agent{SV}_{#1}}
\newcommand{\TM}{\Agent{TM}}
\newcommand{\CM}{\Agent{CM}}
\newcommand{\CC}{\Agent{CC}}

\newcommand{\Sarrive}{\STIMevent{arrive}}
\newcommand{\SmanageA}{\STIMevent{mnge1}}
\newcommand{\SmanageB}{\STIMevent{mnge2}}
\newcommand{\SshipA}{\STIMevent{ship1}}
\newcommand{\SshipB}{\STIMevent{ship2}}
\newcommand{\ScraneA}{\STIMevent{crane1}}
\newcommand{\ScraneB}{\STIMevent{crane2}}
\newcommand{\Sallocate}{\STIMevent{allocd}}
\newcommand{\Sberth}{\STIMevent{berth}}
\newcommand{\Sdock}{\STIMevent{dock}}
\newcommand{\SoperateA}{\STIMevent{oper1}}
\newcommand{\SoperateB}{\STIMevent{oper2}}
\newcommand{\Scarrier}{\STIMevent{carrier}}
\newcommand{\Sassign}{\STIMevent{assgnd}}
\newcommand{\Sserve}{\STIMevent{serve}}
\newcommand{\Supdate}{\STIMevent{served}}
\newcommand{\Sdone}{\STIMevent{done}}
\newcommand{\ScompleteA}{\STIMevent{compl1}}
\newcommand{\ScompleteB}{\STIMevent{compl2}}
\newcommand{\SdepartA}{\STIMevent{deprt1}}
\newcommand{\SdepartB}{\STIMevent{deprt2}}

\newcommand{\PCTstimuli}{$\{\Sarrive$, $\SmanageA$, $\SmanageB$, $\SshipA$, $\SshipB$, $\ScraneA$, $\ScraneB$, $\Sallocate$, $\Sberth$, $\Sdock$, $\SoperateA$, $\SoperateB$, $\Scarrier$, $\Sassign$, $\Sserve$, $\Supdate$, $\Sdone$, $\ScompleteA$, $\ScompleteB$, $\SdepartA$, $\SdepartB\}$}

\newcommand{\Kdepart}{\behave{depart}}
\newcommand{\KclearA}{\behave{clear1}}
\newcommand{\KclearB}{\behave{clear2}}
\newcommand{\Kinit}{\behave{init}}
\newcommand{\KmanA}{\behave{man1}}
\newcommand{\KmanB}{\behave{man2}}
\newcommand{\KtimeS}{\behave{srvT}}
\newcommand{\Kpos}{\behave{posn}}
\newcommand{\Kleave}{\behave{leave}}
\newcommand{\Kcrane}{\behave{cranes}}
\newcommand{\Kplan}{\behave{plan}}
\newcommand{\Kdock}{\behave{dock}}
\newcommand{\Krlse}{\behave{rlse}}
\newcommand{\Kallo}{\behave{allo}}
\newcommand{\Kfree}{\behave{free}}
\newcommand{\Kread}{\behave{read}}
\newcommand{\Kcargo}{\behave{cargo}}
\newcommand{\Kseq}{\behave{seq}}
\newcommand{\Kserve}{\behave{serve}}
\newcommand{\Kupdate}{\behave{updt}}
\newcommand{\Koperate}{\behave{oper}}
\newcommand{\Kavail}{\behave{avail}}
\newcommand{\Kassign}{\behave{assgn}}
\newcommand{\Knear}{\behave{near}}
\newcommand{\Kmove}{\behave{move}}

\newcommand{\PCTbehaviors}{$\{\Kdepart$, $\KclearA$, $\KclearB$, $\Kinit$, $\KmanA$, $\KmanB$, $\KtimeS$, $\Kpos$, $\Kleave$, $\Kcrane$, $\Kplan$, $\Kdock$, $\Krlse$, $\Kallo$, $\Kfree$, $\Kread$, $\Kcargo$, $\Kseq$, $\Kserve$, $\Kupdate$, $\Koperate$, $\Kavail$, $\Kassign$, $\Knear$, $\Kmove\}$}

\usepackage{soul}

\definecolor{darkred}{rgb}{0.75,0.0,0.0}
\definecolor{darkgreen}{rgb}{0.0,0.6,0.0}
\definecolor{darkblue}{rgb}{0.0,0.0,0.6}
\definecolor{darkcyan}{rgb}{0.0,0.6,0.6}
\definecolor{darkmagenta}{rgb}{0.6,0.0,0.6}
\definecolor{darkyellow}{rgb}{0.6,0.6,0.0}
\definecolor{lightred}{rgb}{1.0,0.9,0.9}
\definecolor{lightgreen}{rgb}{0.9,1.0,0.9}
\definecolor{lightblue}{rgb}{0.9,0.9,1.0}
\definecolor{lightcyan}{rgb}{0.8,1.0,1.0}
\definecolor{lightmimagenta}{rgb}{1.0,0.8,1.0}
\definecolor{lightyellow}{rgb}{1.0,1.0,0.8}
\definecolor{paleyellow}{rgb}{1.0,1.0,0.8}
\definecolor{amber}{rgb}{1.0,0.8,0.0}
\definecolor{darkamber}{rgb}{1.0,0.5,0.0}
\definecolor{webgreen}{rgb}{0,0.5,0}
\definecolor{webbrown}{rgb}{0.6,0,0}
\definecolor{grey}{rgb}{0.65,0.65,0.65}
\definecolor{purple}{rgb}{0.4,0,0.75}

\newcommand{\mynote}[2]{
	\ifthenelse{\equal{#1}{0}}{\textcolor{darkamber}{#2}}{}
  	\ifthenelse{\equal{#1}{1}}{\textcolor{darkmagenta}{#2}}{}
  	\ifthenelse{\equal{#1}{2}}{\textcolor{darkcyan}{#2}}{}
  	\ifthenelse{\equal{#1}{3}}{\textcolor{darkgreen}{#2}}{}
}

\newcommand{\todiscuss}[2]{
	\ifthenelse{\equal{#1}{0}}{\textcolor{darkamber}{\textbf{TO DISCUSS}: #2}}{}
  	\ifthenelse{\equal{#1}{1}}{\textcolor{darkmagenta}{\textbf{TO DISCUSS}: #2}}{}
  	\ifthenelse{\equal{#1}{2}}{\textcolor{darkcyan}{\textbf{TO DISCUSS}: #2}}{}
  	\ifthenelse{\equal{#1}{3}}{\textcolor{darkgreen}{\textbf{TO DISCUSS}: #2}}{}
}

\newcommand{\todo}[2]{
	\ifthenelse{\equal{#1}{0}}{\textcolor{darkamber}{\textbf{\underline{TO DO}}: #2}}{}
  	\ifthenelse{\equal{#1}{1}}{\textcolor{darkmagenta}{\textbf{\underline{TO DO}}: #2}}{}
  	\ifthenelse{\equal{#1}{2}}{\textcolor{darkcyan}{\textbf{\underline{TO DO}}: #2}}{}
  	\ifthenelse{\equal{#1}{3}}{\textcolor{darkgreen}{\textbf{\underline{TO DO}}: #2}}{}
}

\newcommand{\newissue}[2]{
	\ifthenelse{\equal{#1}{0}}{\noindent\textcolor{darkamber}{\textit{#2}}\\}{}
  	\ifthenelse{\equal{#1}{1}}{\noindent\textcolor{darkmagenta}{\textit{#2}}\\}{}
  	\ifthenelse{\equal{#1}{2}}{\noindent\textcolor{darkcyan}{\textit{#2}}\\}{}
  	\ifthenelse{\equal{#1}{3}}{\noindent\textcolor{darkgreen}{\textit{#2}}\\}{}
}

\newcommand{\system}{port terminal coordination system\@\xspace}
\newcommand{\SYSTEM}{Port Terminal Coordination System\@\xspace}

\begin{document}
	
\title{Evaluating the Exploitability of Implicit Interactions in Distributed Systems
}

\author{\IEEEauthorblockN{Jason Jaskolka}\\
\IEEEauthorblockA{\textit{Systems and Computer Engineering}\\
\textit{Carleton University}\\
Ottawa, ON Canada\\
jason.jaskolka@carleton.ca}
}

\maketitle

\begin{abstract}
	Implicit interactions refer to those interactions among the components of a system that may be unintended and/or unforeseen by the system designers. As such, they represent cybersecurity vulnerabilities that can be exploited to mount cyber-attacks causing serious and destabilizing system effects. In this paper, we study implicit interactions in distributed systems specified using the algebraic modeling framework known as \CCKA~(\CCKAabbrv). To identify and defend against a range of possible attack scenarios, we develop a new measure of exploitability for implicit interactions to aid in evaluating the threat posed by the existence of such vulnerabilities in system designs for launching cyber-attacks. The presented approach is based on the modeling and analysis of the influence and response of the system agents and their \CCKAabbrv specifications. We also demonstrate the applicability of the proposed approach using a prototype tool that supports the automated analysis. The rigorous, practical techniques presented here enable cybersecurity vulnerabilities in the designs of distributed systems to be more easily identified, assessed, and then mitigated, offering significant improvements to overall system resilience, dependability, and security. 
\end{abstract}

\begin{IEEEkeywords}
	Implicit interactions, \CCKA (\CCKAabbrv), exploitability, attack scenarios, cybersecurity.
\end{IEEEkeywords}

\section{Introduction and Motivation}
\label{sec:introduction}
\IEEEPARstart{I}{mplicit} interactions refer to component interactions within a distributed system that may be unfamiliar, unplanned, or unexpected, and either not visible or not immediately comprehensible by the system designers~\cite{Jaskolka2017aa}. These kinds of interactions have also been referred to as \emph{hidden interactions} in the literature, although it is not necessary that they are intentionally hidden from the view of the system designers. Implicit interactions represent previously unknown linkages among system components. Because system designers are generally unaware of such linkages, they indicate the presence of cybersecurity vulnerabilities that can be exploited by attackers. This can have severe consequences in terms of the system's safety, security, and~reliability.

In previous work~\cite{Jaskolka2017aa,Jaskolka2017ab}, we developed a rigorous and systematic approach for identifying the existence of implicit interactions in distributed systems. The approach involves the specification and analysis of the communication among system components using the \CCKA~(\CCKAabbrv) modeling framework~\cite{Jaskolka2014aa,Jaskolka2015ab}. More specifically, the approach verifies whether each possible interaction in a given system exists as part of a characterization of the intended system interactions resulting from the system design. In any system engineering process, the articulation of the expected behaviour and operation of the system results in a set of intended sequences of communication and interaction among the components of the system. This set of intended interactions is typically derived from the system description and requirements explicitly provided by the system designer. Therefore, any interaction that is found to deviate from this expected or intended behaviour is an implicit interaction.

However, while identifying the existence of these vulnerabilities is a critically important initial step, it is also necessary to examine whether they are likely to manifest in real-world systems~\cite{Jaskolka2017aa}. A natural next step is to determine the ways in which such vulnerabilities can be exploited to mount a cyber-attack in the system. This information is critical in assessing the severity of the vulnerabilities, as well as in determining measures to mitigate the potential that they could be exploited in an attack. 

In this paper, we present an approach for evaluating the exploitability of implicit interactions in distributed system designs. The approach is based on \emph{attack scenario determination}, which looks to find the set of possible ways in which a compromised system agent can exploit a particular implicit interaction to mount a cyber-attack that influences the behaviour of other agents in the system. The attack scenario determination involves an analysis of the set of implicit interactions identified using the technique proposed in~\cite{Jaskolka2017ab} and the~\CCKAabbrv specification of the system. Using the results of the attack scenario determination for an identified implicit interaction, we compute a measure of its exploitability to more accurately assess the threat that it poses to the overall safety, security, and reliability of the system. Thus, the key objective of this paper is to provide a systematic approach for evaluating the ways in which implicit interactions can be used to mount cyber-attacks in a given system, as well as specific guidance on ways to modify system designs to reduce the potential exposure to such attacks. Note that while critically important, in this paper, we are not assessing the potential impact that a cyber-attack resulting from the exploitation of an implicit interaction may have on a given system, but rather the ways in which an attacker may use an implicit interaction to affect the behaviour of the system in order to support the development of methodologies and mechanisms for achieving systems with improved dependability and security.

The rest of this paper is organized as follows.
Section~\ref{sec:related_work} compares and contrasts our proposed approach with related work.
Section~\ref{sec:background} provides the required background for the approach and associated mathematical framework, and 
Section~\ref{sec:example} outlines an illustrative example that will be used to demonstrate the proposed approach throughout this paper. 
Section~\ref{sec:overview} provides a high-level overview of the proposed approach for readers that wish to forgo the technical details provided in Sections~\ref{sec:influence_response}--\ref{sec:exploitability}.
Section~\ref{sec:influence_response} develops the theoretical background required for analyzing the influence and response of agents in a distributed system specified using \CCKAabbrv.
Section~\ref{sec:attack_scenarios} articulates the proposed approach for determining the possible attack scenarios that can exploit implicit interactions.
Section~\ref{sec:exploitability} presents a measure of the exploitability of an implicit interaction.
Section~\ref{sec:discussion} presents a summary of our experimental results in evaluating the exploitability of implicit interactions and provides a discussion of the proposed approach. 
Lastly, Section~\ref{sec:conclusion} concludes and discusses future work.

\section{Related Work}
\label{sec:related_work}
In this section, we compare and contrast our contributions with the existing literature related to assessing the exploitability of cybersecurity vulnerabilities, and studying information flows, dependence, and causality in distributed systems.

\subsection{Threat Modeling and Risk Management Frameworks}
\label{sub:threat_modeling_and_risk_management_frameworks}

The \emph{Common Vulnerability Scoring System} (CVSS)~\cite{Mell2007aa} is largely considered the \emph{de facto} standard for quantifying and assessing the severity and risk of security vulnerabilities in computing systems. CVSS metrics are designed to measure the fundamental characteristics of vulnerabilities that can be used to compute measures of exploitability. 
In CVSS, exploitability is a function of metrics called \emph{access vector}, \emph{access complexity}, and \emph{authentication}. The access vector metric provides a measure of how a vulnerability is exploited (\eg locally or remotely). The access complexity metric provides a measure of the amount of effort that is needed to exploit a vulnerability once a malicious agent has gained access to the system. Lastly, the authentication metric provides a measure of the number of times that a malicious agent needs to authenticate itself to successfully exploit the vulnerability.
A number of additional frameworks for measuring cybersecurity-related vulnerabilities and risks, such as the OCTAVE risk management framework~\cite{Caralli2007aa} and the Microsoft Exploitability Index~\cite{Microsoft2014aa}, have also been proposed. More recently, extensions to CVSS that use stochastic modeling to aid in improving decision making and reducing risk have been proposed (\eg~\cite{Abraham2015aa}).

However, CVSS and its associated exploitability analysis, as well as other similar frameworks, have received much criticism due to their perceived subjectivity and lack of specificity in the ways in which values are measured and assigned~(\eg~\cite{Wang2008ac,Younis2016aa}). By developing a rigorous technique for determining the ways in which a vulnerability can be exploited based on the system specification and design, the approach proposed in this paper, as well as the developed measure of exploitability, avoids this kind of subjectivity and lack of specificity.


\subsection{Attack Surfaces}
\label{sub:attack_surfaces}

Generally speaking, attack surfaces are related to exposures enabling a malicious agent to mount a cyber-attack on a system. Attack surfaces are typically considered along a number of different dimensions including interfaces, channels, protocols, and access rights, among others. The intuition behind analyzing and measuring a system's attack surface is based on the idea that the more extensive and exposed the system's attack surface is, the more opportunity for a malicious agent to conduct an attack~\cite{Howard2005aa}. Therefore, many approaches aim to improve security by reducing the attack surface of the system in question. An important element of this process is understanding the system's attack surface. For example,~\cite{Howard2005aa} proposed a metric called the \emph{Relative Attack Surface Quotient}. This was expanded upon in~\cite{Manadhata2011aa} which additionally proposed considering the damage potential-effort ratios associated with possible attacks. A similar approach was proposed in~\cite{Younis2014aa}. More recently researchers have looked to assess the likelihood of an attack by considering the possibility for individual, coordinated, and concurrent attacks~\cite{Samarji2015aa}. New approaches have also developed more objective metrics derived from attack surface, vulnerability, and exploitation analyses by studying software properties~\cite{Younis2016aa}. 

However, attack surface metrics are often designed to measure the exploitability of an entire system, rather than the exploitability of individual vulnerabilities. By comparison, the proposed approach studies the exploitability of each individually identified vulnerability (\ie each implicit interaction). Consequently, this allows us to obtain information that can help in determining where and how to spend valuable resources to mitigate the most severe, or most exploitable, vulnerabilities.


\subsection{Formal Verification and Test-Based Approaches}
\label{sub:formal_verification_static_analysis_and_testing_approaches}

Many existing approaches for conducting vulnerability and exploitability analyses involve the development high-level models of system components for which security-relevant properties can be formalized and analyzed to verify their satisfaction in the composite system. For example,~\cite{Ramakrishnan2002aa} proposed formal approaches for specifying desired security properties and conducting vulnerability and exploitability analyses. Also,~\cite{Cheung2003aa} proposed the \emph{Correlated Attack Modeling Language} (CAML) to help in automatically identifying cyber-attack~scenarios.

Other approaches look to perform formal verification and analyses on program code. The use of such techniques for specifying and analyzing systems is often highly desirable when developing systems with high standards of safety, security, and reliability~\cite{Nhlabatsi2008aa}. In~\cite{Grieco2013aa}, a symbolic analysis approach that operates on disassembled binary code to identify conditions by which a malicious agent can exploit a ``dangerous path'' was proposed. In~\cite{Shaffer2007aa}, an approach for verifying programs represented in a specialized modeling language using a formal security domain model was presented. The approach aimed to detect execution paths that violated the security properties specified in the domain model. However, the ability to identify and analyze exploitable vulnerabilities at earlier stages of system development was desired. As such,~\cite{Almorsy2013aa} provided an approach that studied system architectures to identify possible scenarios and metrics, similar to those derived for attack surfaces, that can be used to determine which system vulnerabilities may be exploitable. The possible scenarios and security metric signatures were formalized using the Object Constraint Language. This allowed for the development of an approach supporting both metric-based and scenario-based architecture security analysis. Similarly,~\cite{Bistarelli2005aa} provided a constraint-based approach for identifying and mitigating cascading network paths that compromise security. 

Test-based approaches aim to provide a proof-of-concept that a given system vulnerability can be exploited by a malicious system agent. Black-box fuzz testing is a common technique used for this purpose (\eg~\cite{Sparks2007aa}). Other test-based approaches look to use static analysis to find potential vulnerabilities in program code, and then combinations of static and dynamic analyses to uncover execution paths within the code for which an exploit can be automatically generated. This is the idea behind \emph{Automatic Exploit Generation} (AEG)~\cite{Avgerinos2014aa}, which generates evidence that identified security vulnerabilities are exploitable. However, testing-based techniques such as AEG and fuzz testing are not easily scalable and are typically only suited for particular types of systems.

By comparison, the approach proposed in this paper is targeted at analyzing systems at much earlier stages of development. Rather than analyzing program code, we aim to analyze the specification and design of distributed cyber-physical systems at a high-level of abstraction. This can allow for substantial savings in terms of the costs associated with minimizing cybersecurity vulnerabilities, and recovering from the effects of a cyber-attack.


\subsection{Models of Information Flow, Dependence, and Causality}
\label{sub:models_of_information_flow_dependence_and_causality}

Among the most well-known approaches for studying the interactions of components in complex distributed systems and networks has been information flow analysis~\cite{Focardi2003aa}. Many approaches targeted at modeling and analyzing information flow with respect to cybersecurity requirements have been proposed using a variety of formalisms such as state machines (\eg~\cite{Shen2007aa}), Petri nets (\eg~\cite{Varadharajan1990aa}), process algebras (\eg~\cite{Focardi1994aa,Focardi2003aa}), typing systems (\eg~\cite{Hristova2006aa,Volpano1996aa}), and axiomatic approaches (\eg~\cite{Andrews1980aa,Sabri2009aa}). Furthermore, models and notions of causality, such as those proposed in~\cite{Lamport1978aa} and~\cite{Ay2008aa} have also provided foundational approaches for studying the dependence of actions in distributed systems.

Numerous other approaches aimed at formally analyzing and verifying of concurrent systems (\eg~\cite{Lamport1977aa,German1992aa,Siegel2011aa}), as well as the formal verification of dynamic and parametrized systems and networks (\eg~\cite{Abdulla2013aa,Namjoshi2015aa}) have also been prosed. These publications have laid important groundwork for approaches aimed at providing assurances that systems operate as expected as they continue to grow in size and complexity.

Although many formalisms and approaches exist for modeling and studying interactions, information flows, and dependencies among components in distributed systems, we propose an alternative approach meant to aid designers, at early stages of system development, in systematically evaluating the ways in which security vulnerabilities in their designs can be exploited to cause unexpected, and potentially unsafe and insecure, system behaviours. Our approach, based on the \CCKAabbrv modeling framework, provides a different and complementary perspective for studying the interactions of system components and evaluating the exploitability of security vulnerabilities in system designs, than what is offered by existing formalisms and approaches.



\section{Background and Preliminaries}
\label{sec:background}
In this section,
we briefly introduce \CCKA and the required preliminaries related to specifying agents and interactions.

\subsection{\CCKA}
\label{sub:c2ka}

\CCKA~(\CCKAabbrv)~\cite{Jaskolka2014aa,Jaskolka2015ab} is an algebraic framework for specifying the concurrent and communicating behaviour of agents in a distributed system. A \CCKAabbrv is a mathematical system consisting of a \leftSemimodule{\stim}~$\ActSemimodule$ and a \rightSemimodule{\cka}~$\OutSemimodule$ which characterize how a stimulus structure~$\stim$ and a \CKAabbrv~$\cka$ mutually act upon one another to describe the response invoked by a stimulus on an agent behaviour as a next behaviour and a next stimulus.~$\ActSemimodule$ describes how the stimulus structure~$\stim$ acts upon the \CKAabbrv~$\cka$ via the \emph{next behaviour mapping}~$\actOp$ and~$\OutSemimodule$ describes how the \CKAabbrv~$\cka$ acts upon the stimulus structure~$\stim$ via the \emph{next stimulus mapping}~$\outOp$. The formal definition of a \CCKAabbrv is given in Definition~\ref{def:C2KA}. 

\begin{definition}[\CCKAabbrv\ --- \eg~\cite{Jaskolka2015ab}]
\label{def:C2KA}
	A \emph{\CCKAabbrv} is a system~$\CCKAstructure$, where~$\stim = \STIMstructure$ is a stimulus structure and~$\cka = \CKAstructure$ is an atomic \CKAabbrv such that~$\ActSemimodule$ is a unitary and zero-preserving \emph{\leftSemimodule{\stim}} with next behaviour mapping~$\lActSig$ and~$\OutSemimodule$ is a unitary and zero-preserving \emph{\rightSemimodule{\cka}} with next stimulus mapping~$\lOutSig$, where the following axioms are satisfied for all~$a,b,c \in \CKAset$ and~$s,t \in \STIMset$:
	\begin{enumerate}[(a)]
			\item \label{def:cascading_axiom}
				$\lAct{(a \CKAseq b)}{s} = (\lAct{a}{s}) \CKAseq \bigP{\lAct{b}{\lOut{a}{s}}}$
			\item \label{def:cascading_output_axiom}
				$a \CKAle c \Ors b = 1 \Ors (\lAct{a}{s}) \CKAseq \bigP{\lAct{b}{\lOut{c}{s}}} = 0$
			\item \label{def:sequential_output_axiom}
				$\lOut{a}{s \STIMdot t} = \lOutbig{(\lAct{a}{t})}{s} \STIMdot \lOut{a}{t}$
			\item \label{def:idle_agent_law}
				$s = \Dstim \Ors \lAct{1}{s} = 1$
			\item \label{def:neutral_stimulus_law}
				$a = 0 \Ors \lOut{a}{\Nstim} = \Nstim$  
	\end{enumerate}
\end{definition}	

The reader is referred to the Appendix A for a summary of the algebraic structures mentioned in this discussion, and to~\cite{Jaskolka2014aa,Jaskolka2015ab,Jaskolka2016ad} for a full description of \CCKAabbrv.

Many complex distributed systems involve intensive communication and exchange with their environment, which often includes other systems. Consequently, when modeling such systems, the interactions between the system and its environment need to be carefully taken into account~\cite{Kroger2011aa}. \CCKAabbrv allows for the separation of communicating and concurrent behaviour in a system and its environment and for the expression of the influence of stimuli on agent behaviours, thereby providing the capability to model and capture the dynamic behaviour of complex distributed systems by considering these important interactions.

\subsubsection{Atomic Behaviours and Stimuli}
\label{ssub:atomic_behaviours_and_stimuli}

For a commutative monoid~$\monoid{S}{\cdot}{1}$ with~$a,b \in S$, the divisibility relation~$\divRel$ is defined on~$S$ via~$a \divRel b \mIff \lnotation{\exists}{c}{c \in S}{b = a \cdot c}$. An element~$a$ is called a \emph{unit} if~$a \divRel 1$, and a \emph{non-unit} otherwise. A non-unit~$a$ is called an \emph{atom} if~$a = b \cdot c$ implies that either~$b$ or~$c$ is a unit. We say that~$\monoid{S}{\cdot}{1}$ is \emph{atomic} if every non-unit may be factored into atoms in at least one way. Throughout the remainder of this paper, the set of \emph{atomic behaviours} will be denoted by~$\CKAbasic$, and the set of \emph{atomic stimuli} will be denoted by~$\STIMbasic$.
	

\subsubsection{Sub-Behaviours and Sub-Stimuli}
\label{ssub:sub-behaviours_and_sub-stimuli}

In general, every idempotent semiring~$\semiring{S}{+}{\cdot}{0}{1}$ has a natural partial order~$\le$ on~$S$ defined by~$a \le b \mIff a + b = b$. This means that associated with a \CKAabbrv~$\cka = \CKAstructure$, there is an ordering relation~$\CKAle$ related to the semirings upon which~$\cka$ is built representing the sub-behaviour relation. For behaviours~$a,b \in \CKAset$,~$a \CKAle b$ indicates that~$a$ is a \emph{sub-behaviour} of~$b$ if and only if~$a + b = b$. Similarly, associated with a stimulus structure~$\stim = \STIMstructure$ is an ordering relation~$\STIMle$ representing the sub-stimulus relation. For stimuli~$s,t \in \STIMset$,~$s \STIMle t$ indicates that~$s$ is \emph{sub-stimulus} of~$t$ if and only if~$s \STIMplus t = t$. These notions will play an important role in the attack scenario determination in Section~\ref{sub:attack_scenario_determination}. 


\subsubsection{Agents}
\label{ssub:agents}

In this paper, the term \emph{agent} refers to any system, component, or process whose behaviour consists of discrete actions~\cite{Milner1989aa}. We write~$\agent{A}{a}$ where~$\Agent{A}$ is the name given to an agent and~$a \in \CKAset$ is its behaviour. For~$\agent{A}{a}$ and~$\agent{B}{b}$, we write~$\Agent{A+B}$ to denote the agent~$\bigA{a+b}$. In a similar way, we can extend the remaining operators on behaviours of~$\CKAset$ to their corresponding agents. Thus, agents are defined by simply describing their behaviour, and for this reason, we may use the terms agents and behaviours interchangeably.


\subsubsection{Specifying Agents using \CCKAabbrv}
\label{ssub:specifying_agents_using_c2ka}

\CCKAabbrv provides three levels of specification for the behaviour of agents in distributed systems. The \emph{\levelOne of agents} specifies the next behaviour mapping~$\actOp$ and next stimulus mapping~$\outOp$ for each agent. This involves specifying how each atomic stimulus that can be issued acts upon the atomic behaviours that each agent can have in the given system. The \emph{\levelTwo} specifies each agent behaviour as a \CKAabbrv term. This enables the specification of complex agent behaviours without the need to explicitly articulate the dependencies between agent behaviours, or to further refine agent behaviours into state-based specifications. Lastly, the \emph{\levelThree} provides the state-level specification of each agent behaviour. At this level, the concrete programs for each of the \CKAabbrv terms which specify each agent behaviour are given using any suitable programming or specification language. In this paper, we use Dijkstra's guarded command language~\cite{Dijkstra1975aa} for this purpose. 

To determine the possible ways that agents in a distributed system can interact and influence each other's behaviour, we need to consider multiple levels of abstraction in the specification and analysis of the system. These multiple levels of abstraction correspond to the possibility for agents to communicate via stimuli (\ie message-passing communication) by considering the stimulus-response and \levelTwo{s}, or via shared environments (\ie via shared variable communication) by considering the \levelThree{s}. For this reason, all three levels of specification are required for the analysis approach presented in Section~\ref{sec:attack_scenarios}. 



\subsection{Agent Interactions}
\label{sub:agent_interactions}

A distributed system may consist of numerous agents, and may feature complex agent interactions for synchronizing and sequencing behaviour, or coordinating access to shared resources, for example. In a distributed system formed by a set~$\A$ of agents, agent interactions are represented as sequences of agents in the following form:~$\impX{n} \deq \impN$ where each~$\AgentX{i} \in \A$ for all~$0 \le i \le n$, and each~$\typeX{j} \in \set{\stim, \env}$ for all~$1 \le j \le n$. In an agent interaction,~$\toS$ denotes direct communication via stimuli and~$\toE$ denotes direct communication via shared environments. These notions are discussed in more detail in Sections~\ref{sub:influencing_stimuli} and \ref{sub:influencing_variables}. In this way, an interaction~$\impX{n}$ can be written recursively such that~$\impX{1} \deq \AgentX{1} \toX{1} \AgentX{0}$ and~$\impX{n} \deq \AgentX{n} \toX{n} \impX{n-1}$.
The length of an interaction~$\impX{n}$ (denoted~$\abs{\impX{n}}$) is counted by the number of direct communications of which it is comprised (\ie~$\abs{\impX{n}} = n-1$).

Additionally, for an interaction~$\impX{n}$, we call~$\AgentX{n}$ the \emph{source agent} of the interaction,~$\AgentX{0}$ the \emph{sink agent} of the interaction, and each~$\AgentX{i-1}$ the \emph{neighbouring agent} of~$\AgentX{i}$ for all~$1 \le i \le n$. Moreover, when we refer to a \emph{compromised agent}, we mean any agent~$\Agent{A} \in \A$ that can behave in a way that is not consistent with its original or intended specification, meaning that it has the ability to issue any stimulus~$s \in \STIMset$ and/or alter its concrete behaviour (\eg by defining a program variable~$v$ in the set of all program variables in the state space of the system).


\section{An Illustrative Example}
\label{sec:example}
Distributed systems, where each agent is responsible for providing some information, or controlling some element of the system, play a vital role in many critical industries, such as critical infrastructures, aerospace, automotive, and industrial manufacturing. In this paper, we consider an illustrative maritime port container terminal coordination system adapted from~\cite{Henesey2006aa}.

\subsection{\SYSTEM Description}
\label{sub:system_description}

The \system consists of six classes of agents, which consist of both cyber components (\eg control software) and physical components. The \emph{\portcaptain~$(\PC)$} decides how the ship will be managed once it arrives at the port, and initializes the arriving ship's information. The \emph{\shipmanager{s}~$(\SM{i})$} are assigned to manage the loading/unloading of an arriving ship, and for determining its desired service time. For each \shipmanager, there is an associated \emph{\stevedore~$(\SV{i})$} that is responsible for requesting cranes and planning the loading/unloading operations for an arriving ship. The \emph{\terminalmanager~$(\TM)$} allocates the berth points (\ie the docking position) for an arriving ship, and allocates cranes to service a ship based on requests from the \stevedore{s}. The \crane~$(\CM)$ is responsible for coordinating the cranes to carry out the loading/unloading sequence as efficiently as possible. The \carrier~$(\CC)$ is responsible for managing the straddle carrier positions to aid in loading/unloading the ship, and moving containers on the ground of the shipping yard. 

When a ship approaches the port, it transmits an~$\Sarrive$ message to begin the port-side operations. The \portcaptain~$\PC$ responds to the~$\Sarrive$ message by determining how to manage the ship upon its arrival at the port, and initializing its record of the ship's information. For the purpose of illustration in this paper, assume that there are two \shipmanager{s}~$\SM{1}$ and~$\SM{2}$. This means that there are also two \stevedore{s}~$\SV{1}$ and~$\SV{2}$ assigned to their corresponding \shipmanager{s}. Assume that~$\PC$ non-deterministically chooses to use~$\SM{1}$ and~$\SV{1}$, or~$\SM{2}$ and~$\SV{2}$ to manage arriving ships. Based on its choice,~$\PC$ sends either a~$\SmanageA$ message or a~$\SmanageB$ message to notify the appropriate \shipmanager that is has been assigned to manage the arriving vessel. In what follows, we describe the operation of the system in the case where a~$\SmanageA$ message is sent. An analogous operation involving~$\SM{2}$ and~$\SV{2}$ occurs in the event that~$\PC$ sends a~$\SmanageB$ message. 

Upon receiving a~$\SmanageA$ message, \shipmanager~$\SM{1}$ responds by reading the ship information from~$\PC$ and computing the desired service time according to the following formula:~$t_{\text{service}} = t_{\text{depart}} - t_{\text{arrive}} - t_{\text{wait}}$. After computing the desired service time,~$\SM{1}$ issues a~$\SshipA$ request. Upon receiving a~$\SshipA$ request, \stevedore~$\SV{1}$ tries to satisfy the request by calculating the number of cranes~$n$ needed to service the ship. This calculation is based on the number of containers~$c$, the desired service time~$t_{\text{service}}$, the average efficiency of the cranes~$x$ (moves per hour), and according to the following formula:~$n = c / (x * t_{\text{service}})$. Once completed,~$\SV{1}$ sends a~$\ScraneA$ request. 

The \terminalmanager~$\TM$ responds to the~$\ScraneA$ request by allocating the berth points for the ship and then allocating cranes to service the ship. After completing the allocations,~$\TM$ issues an~$\Sallocate$ message. The waiting~$\SV{1}$ responds to the~$\Sallocate$ message by determining a ship bay allocation plan based on the crane allocation and berth position. Upon completion of the plan,~$\SV{1}$ conveys the berth position to~$\SM{1}$ via a~$\Sberth$ message.~$\SM{1}$ responds to the~$\Sberth$ message by updating the ship docking position, and issuing a~$\Sdock$ command to notify the ship that it may proceed to dock at the given berth position.~$\SV{1}$ also responds to the~$\Sdock$ command by recording that the ship is docked and issuing a~$\SoperateA$ command. 

The \crane~$\CM$ responds to the~$\SoperateA$ command by reading the ship bay plan, determining the containers that should be loaded/unloaded to each bay, and issuing a~$\Scarrier$ request. The \carrier~$\CC$ responds to the~$\Scarrier$ request by determining the availability of the straddle carriers, assigning a set of them to service the crane, and issuing an~$\Sassign$ message when completed.~$\CM$ responds to the~$\Sassign$ message by determining the loading/unloading sequence and the position to which a straddle carrier should be moved. Once completed,~$\CM$ issues a~$\Sserve$ request.~$\CC$ responds to the~$\Sserve$ request by determining the closest assigned straddle carrier and moving it to the requested position. 

\newcommand{\Smess}[4][0]{
	\stepcounter{seqlevel}
	\path
	(#2)+(0,-\theseqlevel*\unitfactor-0.7*\unitfactor) node (mess from) {};
	\addtocounter{seqlevel}{#1}
	\path
	(#4)+(0,-\theseqlevel*\unitfactor-0.7*\unitfactor) node (mess to) {};
	\draw[->,>=angle 60] (mess from) -- (mess to) node[midway, above]
	{#3};

	\node (#4 from) at (mess from) {};
	\node (#4 to) at (mess to) {};
}
\newcommand{\SmessPAR}[4][0]{
	\path
	(#2)+(0,-\theseqlevel*\unitfactor-0.7*\unitfactor) node (mess from) {};
	\addtocounter{seqlevel}{#1}
	\path
	(#4)+(0,-\theseqlevel*\unitfactor-0.7*\unitfactor) node (mess to) {};
	\draw[->,>=angle 60] (mess from) -- (mess to) node[midway, above]
	{#3};

	\node (#4 from) at (mess from) {};
	\node (#4 to) at (mess to) {};
}
\newcommand{\SmessX}[5][0]{
	\stepcounter{seqlevel}
	\path
	(#2)+(0,-\theseqlevel*\unitfactor-0.7*\unitfactor) node (mess from) {};
	\addtocounter{seqlevel}{#1}
	\path
	(#4)+(0,-\theseqlevel*\unitfactor-0.7*\unitfactor) node (mess to) {};
	\draw[->,>=angle 60] (mess from) -- (mess to) node[midway, above, xshift=#5em]
	{#3};

	\node (#4 from) at (mess from) {};
	\node (#4 to) at (mess to) {};
}
\newcommand{\SmessPARX}[5][0]{
	\path
	(#2)+(0,-\theseqlevel*\unitfactor-0.7*\unitfactor) node (mess from) {};
	\addtocounter{seqlevel}{#1}
	\path
	(#4)+(0,-\theseqlevel*\unitfactor-0.7*\unitfactor) node (mess to) {};
	\draw[->,>=angle 60] (mess from) -- (mess to) node[midway, above, xshift=#5em]
	{#3};

	\node (#4 from) at (mess from) {};
	\node (#4 to) at (mess to) {};
}
\newcommand{\Emess}[4][0]{
	\stepcounter{seqlevel}
	\path
	(#2)+(0,-\theseqlevel*\unitfactor-0.7*\unitfactor) node (mess from) {};
	\addtocounter{seqlevel}{#1}
	\path
	(#4)+(0,-\theseqlevel*\unitfactor-0.7*\unitfactor) node (mess to) {};
	\draw[dashed,->,>=angle 60] (mess from) -- (mess to) node[midway, above]
	{\var{#3}};

	\node (#3 from) at (mess from) {};
	\node (#3 to) at (mess to) {};
}
\newcommand{\EmessPAR}[4][0]{
	\path
	(#2)+(0,-\theseqlevel*\unitfactor-0.7*\unitfactor) node (mess from) {};
	\addtocounter{seqlevel}{#1}
	\path
	(#4)+(0,-\theseqlevel*\unitfactor-0.7*\unitfactor) node (mess to) {};
	\draw[dashed,->,>=angle 60] (mess from) -- (mess to) node[midway, above]
	{\var{#3}};

	\node (#3 from) at (mess from) {};
	\node (#3 to) at (mess to) {};
}
\newcommand{\EmessX}[5][0]{
	\stepcounter{seqlevel}
	\path
	(#2)+(0,-\theseqlevel*\unitfactor-0.7*\unitfactor) node (mess from) {};
	\addtocounter{seqlevel}{#1}
	\path
	(#4)+(0,-\theseqlevel*\unitfactor-0.7*\unitfactor) node (mess to) {};
	\draw[dashed,->,>=angle 60] (mess from) -- (mess to) node[midway, above, xshift=#5em]
	{\var{#3}};

	\node (#3 from) at (mess from) {};
	\node (#3 to) at (mess to) {};
}
\newcommand{\EmessPARX}[5][0]{
	\path
	(#2)+(0,-\theseqlevel*\unitfactor-0.7*\unitfactor) node (mess from) {};
	\addtocounter{seqlevel}{#1}
	\path
	(#4)+(0,-\theseqlevel*\unitfactor-0.7*\unitfactor) node (mess to) {};
	\draw[dashed,->,>=angle 60] (mess from) -- (mess to) node[midway, above, xshift=#5em]
	{\var{#3}};

	\node (#3 from) at (mess from) {};
	\node (#3 to) at (mess to) {};
}

\begin{figure*}
\centering
\resizebox{\textwidth}{!}{	
	\begin{sequencediagram}

		\newthread[white]{PC}{\parbox[c]{3cm}{\centering \portcaptain\\($\PC$)}}
		\newthread[white]{SMa}{\parbox[c]{3cm}{\centering \shipmanager\\($\SM{1}$)}}
		\newthread[white]{SVa}{\parbox[c]{3cm}{\centering \stevedore\\($\SV{1}$)}}
		\newthread[white]{SMb}{\parbox[c]{3cm}{\centering \shipmanager\\($\SM{2}$)}}
		\newthread[white]{SVb}{\parbox[c]{3cm}{\centering \stevedore\\($\SV{2}$)}}
		\newthread[white]{TM}{\parbox[c]{3cm}{\centering \terminalmanager\\($\TM$)}}
		\newthread[white]{CM}{\parbox[c]{3cm}{\centering \crane\\($\CM$)}}
		\newthread[white]{CC}{\parbox[c]{3cm}{\centering \carrier\\($\CC$)}}

		\tikzstyle{init}=[rectangle, draw=white, minimum width=0em]
			\node[init, left of=PC, node distance=8em] 	(SHIP) {};

		\Smess{SHIP}{$\Sarrive$}{PC}
		
		
		\Smess{PC}{$\SmanageA$}{SMa}
		\Emess{PC}{shipInfo}{SMa}
		\Smess{SMa}{$\SshipA$}{SVa}
		\EmessX{PC}{shipInfo}{SVa}{-3.75}
		\Smess{SVa}{$\ScraneA$}{TM}
		\Emess{SVa}{cranes}{TM}
		\SmessX{TM}{$\Sallocate$}{SVa}{-7.5}
		\SmessPAR{TM}{$\Sallocate$}{SVb}
		\Emess{TM}{berth}{SVa}
		\EmessPARX{PC}{shipInfo}{SVa}{-3.75}
		\Smess{SVa}{$\Sberth$}{SMa}
		\SmessPAR{SVa}{$\Sberth$}{SMb}
		\Emess{SVa}{berthPos}{SMa}
		\Smess{SMa}{$\Sdock$}{SVa}
		\SmessPARX{SMa}{$\Sdock$}{SHIP}{-3.75}
		\SmessPAR{SMa}{$\Sdock$}{SVb}
		\SmessX{SVa}{$\SoperateA$}{CM}{3.75}		
		\EmessX{SVa}{bayPlan}{CM}{3.75}
		\Smess{CM}{$\Scarrier$}{CC}		
		\Emess{CM}{containers}{CC}		
		\Smess{CC}{$\Sassign$}{CM}		
		\Emess{CC}{carrierAssign}{CM}	
		\Smess{CM}{$\Sserve$}{CC}		
		\Emess{CM}{position}{CC}		
		\Smess{CC}{$\Supdate$}{CM}		
		\Emess{CC}{carrierState}{CM}
		\SmessX{CM}{$\Sdone$}{SVa}{-15.25}	
		\SmessPARX{CM}{$\Sdone$}{SVb}{3.75}	
		\Smess{SVa}{$\ScompleteA$}{TM}	
		\SmessPAR{SVa}{$\ScompleteA$}{SMa}
		\Smess{SMa}{$\SdepartA$}{PC}	
		\SmessPARX{SMa}{$\SdepartA$}{SHIP}{-3.75}

%



	\end{sequencediagram}}
	\caption{Intended interactions for the \system when using \shipmanager~$\SM{1}$.}
	\label{fig:port_container_terminal}	
	\vspace{-1em}
\end{figure*}	

Once completed,~$\CC$ issues a~$\Supdate$ message.~$\CM$ responds to the~$\Supdate$ message by updating the ship bay plan and carrying out its operations (load/unload). After doing so,~$\CM$ issues a~$\Sdone$ message. The waiting~$\SV{1}$ responds to the~$\Sdone$ message by reseting its record of the ship information and issuing a~$\ScompleteA$ message. Upon receiving a~$\ScompleteA$ message,~$\Agent{TM}$ frees the berth and crane allocations. Also in response to the~$\ScompleteA$ message,~$\SM{1}$ resets its record of the ship information and sends a~$\SdepartA$ message.~$\PC$ responds to the~$\SdepartA$ message by freeing the ship manager allocation and its record of the ship information.
		   
The operation of the port container terminal when using \shipmanager~$\SM{1}$ can be visualized as shown in the sequence diagram given in Fig.~\ref{fig:port_container_terminal}, where the solid arrows denote message-passing communication (\ie communication via stimuli) and the dashed arrows denote shared variable communication (\ie communication via shared environments). 



\subsection{\CCKAabbrv Specification of the System}
\label{sub:c2ka_specification}


To specify the given \system described in Section~\ref{sub:system_description} using \CCKAabbrv, we first identify the set of system agents, namely the set~$\A$ consisting of the agents:~$\{\PC$, $\SM{1}$, $\SM{2}$, $\SV{1}$, $\SV{2}$, $\TM$, $\CM$, $\CC\}$. Next, we identify the set of atomic stimuli that can be issued by the system agents and the set of atomic behaviours that the agents can exhibit. These sets are derived from the system description, and are used to generate the support sets of the stimulus structure~$\stim$ and the \CKAabbrv~$\cka$ that comprise the \CCKAabbrv to be used for the specification. For the \system, the set~$\STIMset$ is generated using the operations of stimulus structures and the set of atomic stimuli~\PCTstimuli. Similarly, the set~$\CKAset$ is generated using the operations of CKA and the set of atomic behaviours~\PCTbehaviors. Lastly, using the constructed \CCKAabbrv, we develop the three levels of specification (see Section~\ref{ssub:specifying_agents_using_c2ka}) for each agent in the system. 

Using the \CCKAabbrv constructed above, the \levelOne{s} of the \system agents are compactly specified. The specification shown below depicts the \levelOne of the \shipmanager~$\SM{1}$:

\scalebox{0.9}{\parbox{\linewidth}{
\begin{equation*}
	\hspace{-1.5em}
	\begin{array}{rclcrcl}
		\lAct{\KtimeS}{\Sberth} &=& \Kpos &\quad&
		\lOut{\KtimeS}{\Sberth} &=& \Sdock\\
		\lAct{\Kpos}{\ScompleteA} &=& \Kleave&\quad&
		\lOut{\Kpos}{\ScompleteA} &=& \SdepartA\\
		\lAct{\Kleave}{\SmanageA} &=& \KtimeS&\quad&
		\lOut{\Kleave}{\SmanageA} &=& \SshipA
	\end{array}
\end{equation*}}}
where $\lAct{a}{s} = a$ and $\lOut{a}{s} = \Nstim$ for all other~$a \in \CKAset$ and~$s \in \STIMset$ in the \CCKAabbrv constructed above. Analogous specifications can be derived for the remaining system agents and can be found in Appendix C.


Based on the description of the system from Section~\ref{sub:system_description}, the \levelTwo of each agent is derived and shown in Fig.~\ref{fig:abstract_behaviours}.

\begin{figure}[t]
	\centering
	\hspace{-3em}
	\scalebox{0.85}{\parbox{\linewidth}{
	\begin{eqnarray*}
		\PC 	&\mapsto& \bigA{(\KmanA + \KmanB) \CKAseq \Kinit + (\KclearA + \KclearB) \CKAseq \Kdepart} 	\\
		\SM{i} 	&\mapsto& \bigA{\KtimeS + \Kpos + \Kleave}												\\
		\SV{i} 	&\mapsto& \bigA{\Kcrane + \Kplan + \Kdock + \Krlse}									\\
		\TM 	&\mapsto& \bigA{\Kallo + \Kfree}															\\
		\CM 	&\mapsto& \bigA{\Kread \CKAseq \Kcargo + \Kseq \CKAseq \Kserve + \Kupdate \CKAseq \Koperate}	\\
		\CC 	&\mapsto& \bigA{\Kavail \CKAseq \Kassign + \Knear \CKAseq \Kmove}
	\end{eqnarray*}}}
	\caption{Abstract behaviour specification of the \system agents.}
	\label{fig:abstract_behaviours}
\end{figure}

Finally, we use a fragment of Dijkstra's guarded command language~\cite{Dijkstra1975aa} to provide the \levelThree of each system agent. This involves specifying the concrete programs corresponding to the \levelTwo of each agent. Fig.~\ref{fig:agent_SMa} depicts the \levelThree of the \shipmanager~$\SM{1}$. Once again, analogous specifications can be derived for the remaining system agents and can be found in Appendix C.

\begin{figure}[ht!]
	\centering
	\hspace{-6em}
	\scalebox{0.775}{\parbox{\linewidth}{
	\begin{eqnarray*}
		\KtimeS		&\deq& 	\varL{serviceT[1]} := \varL{departT[1]} - \varL{arriveT[1]} - \varL{waitT[1]}	\\
		\Kpos		&\deq& 	\varL{dockPos[1]} := \varL{berthPos[1]}										\\
		\Kleave	&\deq& 	\varL{dockPos[1]} := \nulls; \varL{serviceT[1]} := 0							
	\end{eqnarray*}}}
	\caption{Concrete behaviour specification of the \shipmanager~$\SM{1}$ behaviours.}
	\label{fig:agent_SMa}
	\vspace{-1em}
\end{figure}



\subsection{Implicit Interactions Present in the System}
\label{sub:implicit_interactions}

Given the specification of the \system, we can identify the implicit interactions that are present in the system using the approaches presented in~\cite{Jaskolka2017aa,Jaskolka2017ab}. After performing a full system analysis, it can be shown that there are 3902 implicit interactions out of the 4596 total possible system interactions. In particular, the results of the analysis show that there are 19 implicit interactions from the \stevedore~$\SV{1}$ to the \stevedore~$\SV{2}$, which are shown in Fig.~\ref{fig:implicit}.

\begin{figure}[t]
	\centering 
	\resizebox{0.75\linewidth}{!}{
	\begin{minipage}[t]{\linewidth}
		\begin{eqnarray*}
 			p_1		&\deq& \SV{1} \toE \CM \toS \SV{2}							\\
 			p_2		&\deq& \SV{1} \toS \CM \toS \SV{2}							\\
 			p_3		&\deq& \SV{1} \toS \SM{2} \toS \PC \toE \SM{1} \toS \SV{2}	\\
 			p_4		&\deq& \SV{1} \toS \SM{2} \toS \PC \toS \SM{1} \toS \SV{2}	\\
 			p_5		&\deq& \SV{1} \toE \SM{1} \toS \PC \toE \SV{2}				\\
 			p_6		&\deq& \SV{1} \toS \SM{1} \toS \PC \toE \SV{2}				\\
 			p_7		&\deq& \SV{1} \toS \SM{2} \toS \PC \toE \SV{2}				\\
 			p_8		&\deq& \SV{1} \toS \SM{2} \toE \SV{2}						\\
 			p_9		&\deq& \SV{1} \toE \SM{1} \toS \PC \toE \SM{2} \toE \SV{2}	\\
 			p_{10}	&\deq& \SV{1} \toS \SM{1} \toS \PC \toE \SM{2} \toE \SV{2}	\\
 			p_{11}	&\deq& \SV{1} \toE \SM{1} \toS \PC \toS \SM{2} \toE \SV{2}	\\
 			p_{12}	&\deq& \SV{1} \toS \SM{1} \toS \PC \toS \SM{2} \toE \SV{2}	\\
 			p_{13}	&\deq& \SV{1} \toE \TM \toE \SV{2}							\\
 			p_{14}	&\deq& \SV{1} \toS \TM \toE \SV{2}							\\
 			p_{15}	&\deq& \SV{1} \toS \SM{2} \toS \SV{2}						\\
 			p_{16}	&\deq& \SV{1} \toE \SM{1} \toS \PC \toE \SM{2} \toS \SV{2}	\\
 			p_{17}	&\deq& \SV{1} \toS \SM{1} \toS \PC \toE \SM{2} \toS \SV{2}	\\
 			p_{18}	&\deq& \SV{1} \toE \SM{1} \toS \PC \toS \SM{2} \toS \SV{2}	\\
 			p_{19}	&\deq& \SV{1} \toS \SM{1} \toS \PC \toS \SM{2} \toS \SV{2}	\\
		\end{eqnarray*}
  	\end{minipage}}
	\caption{Identified implicit interactions from~$\SV{1}$ to~$\SV{2}$ in the \system.}	
	\label{fig:implicit}
\end{figure}

Implicit interactions in a system are possible due to the potential for out-of-sequence reads from and/or writes to shared variables, and/or the potential for out-of-sequence stimuli to be issued by system agents. This kind of unexpected behaviour could be the result of agents experiencing some kind of compromise or failure. As an example, consider the implicit interaction represented as~$p_7 \deq \SV{1} \toS \SM{2} \toS \PC \toE \SV{2}$. The existence of this implicit interaction indicates that it is possible for a compromised~$\SV{1}$ to influence the behaviour of~$\SV{2}$ indirectly via~$\SM{2}$ and~$\PC$.


This \system will serve as a running example throughout the remainder of this paper to demonstrate the approach for determining the possible attack scenarios, and for evaluating the exploitability of the subset of implicit interactions shown in Fig.~\ref{fig:implicit} that are present in the system. While we present this example within the context of maritime port operations, the proposed approaches are applicable in nearly all distributed systems.


\section{Overview of the Proposed Approach}
\label{sec:overview}
This section provides a high-level overview of the proposed approach for determining the possible attack scenarios and assessing the exploitability of implicit interactions. It presents the main ideas of the proposed approach for readers that wish to skip the technical details presented in Sections~\ref{sec:influence_response}--\ref{sec:exploitability}.

\subsection{Highlights of the Proposed Approach}
\label{sub:the_proposed_approach}

Determining how system agents are capable of influencing each other's behaviour is an important part of uncovering how an implicit interaction can be exploited to mount a cyber-attack. To make this determination, we first need to study the potential for communication via stimuli and via shared environments in a system specified using \CCKAabbrv. As an example, consider the \shipmanager~$\SM{1}$ from the port container terminal example described in Section~\ref{sec:example}. To determine how~$\SM{1}$ can be influenced from the perspective of message-passing communication, we need to study the potential for direct communication via stimuli among system agents to determine the set of stimuli that will cause an observable change in the behaviour of~$\SM{1}$. Similarly, to determine how~$\SM{1}$ can be influenced from the perspective of shared variable communication, we need to study the potential for direct communication via shared environments to determine the set of program variables that are referenced in the concrete behaviour of~$\SM{1}$. In each of these cases, we are looking to capture the ways in which a compromised agent within the system can directly influence the behaviour of the agent under consideration. The formulation of these notions are provided in Section~\ref{sec:influence_response}.

Once we have determined how a system agent can be directly influenced by other agents in a given system, we need to extend these ideas to determine how an agent can be indirectly influenced by studying the pattern of communication dictated by a particular system interaction. Generally speaking, we are looking to find the stimuli or program variables that a source agent can use to influence its neighbouring agent in the interaction so that the neighbouring agent issues a stimulus or defines a program variable that influences the behaviour of its neighbouring agent, and so on until the behaviour of the sink agent is influenced in some way. The idea is to identify the set of possible scenarios that will cause a kind of ``chain-reaction'' through the given implicit interaction. As an example, consider the implicit interaction represented as $p_7 \deq \SV{1} \toS \SM{2} \toS \PC \toE \SV{2}$ that was identified in the \system in Section~\ref{sec:example}. Since the interaction dictates that~$\SV{1}$ interacts with~$\SM{2}$ via stimuli (\ie $\SV{1} \toS \SM{2}$), we need to determine the set of stimuli that can be issued by a compromised~$\SV{1}$ to influence the behaviour of~$\SM{2}$. However, we must ensure that the resulting behaviour of~$\SM{2}$, when influenced by~$\SV{1}$, will lead to the issuance of a stimulus that can influence the behaviour of~$\PC$, since, according to the given interaction,~$\SM{2}$ interacts with~$\PC$ via stimuli (\ie $\SM{2} \toS \PC$). Again, since the interaction shows that~$\PC$ interacts with~$\SV{2}$ via shared environments (\ie $\PC \toE \SV{2}$), it must be ensured that the resulting behaviour of~$\PC$ defines a program variable that is referenced by~$\SV{2}$. Only in this way, will the actions of the compromised~$\SV{1}$ ultimately influence the behaviour of~$\SV{2}$, and will the implicit interaction be exploitable. The technical details and formulation of this approach, which we call \emph{attack scenario determination}, can be found in Section~\ref{sec:attack_scenarios}.

After having determined the set of possible attack scenarios which allow for an implicit interaction to be exploited in a given system, we would like to have a measure of the overall exploitability of the given implicit interaction so that we can better assess the threat that it may pose to the system. Using the results of the attack scenario determination, we devise a new measure of severity for an implicit interaction called \emph{exploitability}. The derivation of the exploitability measure can be found in Section~\ref{sec:exploitability}. This new measure of severity provides a way to compare implicit interactions, as well as actionable information for system designers to determine where and how to spend valuable resources in mitigating the most severe vulnerabilities that exist in their designs. This insight can offer significant improvements to the overall the safety, security, and reliability of the system.

As mentioned above, the complete technical details of what has been presented in this section can be found in Sections~\ref{sec:influence_response}--\ref{sec:exploitability}. Readers that wish to forgo those details may skip ahead to Section~\ref{sec:discussion} for a summary of our experimental results and a discussion of the proposed approach.


\subsection{Tool Support}
\label{sub:tool_support}

We use a prototype software tool to support the automated analysis of the exploitability of implicit interactions present in distributed systems specified using \CCKAabbrv.  It allows for the specification of systems using \CCKAabbrv and automatically identifies the implicit interactions in a given system. For this paper, the tool has been extended to also automatically determine the potential attack scenarios and compute the exploitability measure for identified implicit interactions. The tool is implemented in \emph{Haskell} and uses the \emph{Maude} term rewriting system~\cite{Clavel2003aa}.


\section{Analyzing the Influence and Response of System Agents}
\label{sec:influence_response}
In this section, we formulate and capture the ways in which an agent can be influenced by another agent by studying the communication via stimuli and via shared environments in a system specified using \CCKAabbrv. Given an implicit interaction identified to exist in a given system, we are interested in determining the exact stimuli or program variables that a compromised source agent of the interaction can use to ultimately influence the behaviour of the sink agent of the interaction thereby causing the system to experience unintended or unanticipated behaviours. The formulations developed in this section will be applied in Section~\ref{sec:attack_scenarios} to determine the possible attack scenarios for a given implicit interaction. In the following discussion, we consider a distributed system formed by a set~$\A$ of agents with agents~$\Agent{A},\Agent{B} \in \A$ such that~$\Agent{A} \neq \Agent{B}$.

\subsection{Influencing Stimuli}
\label{sub:influencing_stimuli}

In a distributed system, stimuli are required to initiate agent behaviours. As such, we assume that an agent needs to be influenced by a stimulus before it can issue a stimulus that might influence the behaviour of another agent. Formally, this assumption is articulated as~$(\lAct{a}{s} = a) \mImp (\lOut{a}{s} = \Nstim)$ for all~$s \in \STIMset\STdiff\set{\Dstim}$ and~$a \in \CKAset\STdiff\set{0}$. 

This assumption is motivated by the fact that, in order for an agent to generate and issue a stimulus (\ie to send a message or signal), a concrete program needs to be executed. As an analogy, consider a system to be an arrangement of dominoes where each domino represents a system agent. For a domino to fall over, and potentially cause other dominoes to fall over, some kind of stimulus (\eg a push) is required; a domino cannot simply fall over by itself. This means that an agent needs some kind of external influence to initiate its behaviour and execute its programs before it can influence any other agent(s) in the system. Such external influences may result from systems outside the boundaries of the system being considered. 

Throughout this paper, when we consider a distributed system and its constituent agents, each agent is subjected to each stimulus that is presented to the system, and every stimulus invokes a \emph{response} from an agent. When the behaviour of an agent changes as a result of the response, we say that the stimulus \emph{influences} the behaviour of the agent.

We say that~$\agent{A}{a}$ has the \emph{potential for direct communication via stimuli} with~$\agent{B}{b}$ (denoted by~$\STIMcommD{\Agent{A}}{\Agent{B}}$) if and only if~$\biglnotation{\exists}{s,t}{s,t \in \STIMbasic \nAnd t \STIMle \lOut{a}{s}}{\lAct{b}{t} \neq b}$ where~$\STIMbasic$ is the set of all atomic stimuli~\cite{Jaskolka2014ac}. This means that if there exists an atomic sub-stimulus that is generated by~$\Agent{A}$ that causes an observable change in the behaviour of~$\Agent{B}$, then there is a potential for direct communication via stimuli from~$\Agent{A}$ to~$\Agent{B}$. In this way, we can alternatively say that if~$\STIMcommD{\Agent{A}}{\Agent{B}}$, then there is at least one way in which~$\Agent{A}$ can \emph{influence} the behaviour of~$\Agent{B}$ by communication via stimuli, and that is by issuing the atomic stimulus~$t$. 

However, we are interested in determining all of the possible ways in which the behaviour of an agent can be influenced by another agent in the system. By relating the notion of influence to the formal definition of the potential for direct communication via stimuli, we determine the set of stimuli that can influence the behaviour of a given agent.

\begin{definition}[Influencing Stimuli]
\label{def:influencing_stimuli}
	Let~$\CCKAstructure$ be a \CCKAabbrv. The \emph{influencing stimuli} of an agent~$\agent{A}{a}$ with~$a \in \CKAset$ is the set given by: $\infl{\Agent{A}} = \sets{s \in \STIMbasic}{\lAct{a}{s} \neq a}$.
\end{definition}

The \emph{influencing stimuli} of~$\Agent{A}$ is the set of all atomic stimuli that cause an observable change in the behaviour of~$\Agent{A}$. The influencing stimuli of an agent can be used to determine how other agents in the system can directly influence the given agent's behaviour. For instance, to directly influence the behaviour of~$\Agent{A}$, an agent must issue some stimulus~$s \in \infl{\Agent{A}}$.

Given a \CCKAabbrv, it should be noted that since~$\ActSemimodule$ is zero-preserving, every agent behaviour becomes inactive when subjected to the deactivation stimulus~$\Dstim$ (\ie~$\lAct{a}{\Dstim} = 0$ for all~$a \in \CKAset$). This means that every agent, other than the inactive agent~$0$, can be influenced by the deactivation stimulus~$\Dstim$. Similarly, since~$\ActSemimodule$ is unitary, every agent behaviour remains unchanged by the neutral stimulus~$\Nstim$ (\ie~$\lAct{a}{\Nstim} = a$ for all~$a \in \CKAset$). This means that the neutral stimulus~$\Nstim$ does not influence the behaviour of any agent. For these reasons, we exclude these trivial cases when discussing influencing stimuli in the remainder of this paper.
 
\pagebreak
\begin{example}[Computing the Influencing Stimuli of an Agent]
	Consider the \system described in Section~\ref{sec:example}. The set of influencing stimuli for the \shipmanager~$\SM{1}$ is given by~$\infl{\SM{1}} = \{\Sberth$,~$\ScompleteA$,~$\SmanageA\}$. This means any agent that issues any of the stimuli~$\Sberth$,~$\ScompleteA$, or~$\SmanageA$ will influence the behaviour of~$\SM{1}$.
	When given the system specification, the computation of the set of influencing stimuli for a given system agent can be computed automatically using our prototype software tool.
\end{example}

An agent~$\agent{A}{a}$ is said to have a \emph{fixed point behaviour} if~$\lnotation{\forall}{s}{s \in \STIMset \STdiff \set{\Dstim}}{\lAct{a}{s} = a}$~\cite{Jaskolka2015ab}. This means that a fixed point behaviour is one that remains unchanged in response to all stimuli other than the deactivation stimulus~$\Dstim$. Proposition~\ref{prop:infl_fixed_point} shows how an agent with a fixed point behaviour does not have any non-trivial influencing stimuli. 

\begin{proposition}
\label{prop:infl_fixed_point}
	If an agent~$\agent{A}{a}$ has a fixed point behaviour then~$\infl{\Agent{A}} = \STbot$.

	\begin{proof}
		The proof is straightforward from the definition of a fixed point behaviour.
		The detailed proof can be found in Appendix~B.
	\end{proof}
\end{proposition}

As a direct result of Proposition~\ref{prop:infl_fixed_point}, we have~$\infl{1} = \STbot$ and~$\infl{0} = \STbot$. This means that the idle agent~$1$ and inactive agent~$0$ cannot be influenced by any non-trivial~stimuli.


\subsection{Influencing Variables}
\label{sub:influencing_variables}

We say that~$\agent{A}{a}$ has the \emph{potential for direct communication via shared environments} with~$\agent{B}{b}$ (denoted by~$\ENVcommD{\Agent{A}}{\Agent{B}}$) if and only if~$\dep{b}{a}$ where~$\depOp$ is a \emph{dependence relation}~\cite{Jaskolka2014ac}. In this paper, due to the use of Dijkstra's guarded command language for the \levelThree{s} of agents, such a dependence relation is considered to be a \emph{definition-reference relation} between program variables. Thus if there exists a program variable that is defined in the concrete behaviour specification of~$\Agent{A}$, and referenced in the concrete behaviour specification of~$\Agent{B}$, then there is a potential for direct communication via shared environments from~$\Agent{A}$ to~$\Agent{B}$. 

In what follows, let~$\defV{\Agent{A}}$ and~$\refV{\Agent{A}}$ represent the sets of program variables that are defined and referenced in the \levelThree of agent~$\Agent{A}$, respectively. These sets can be defined by structural induction on programs specified with Dijkstra's guarded command language. We are interested in determining all of the possible ways in which the behaviour of an agent can be influenced by another agent in the system. By relating the notion of influence to the formal definition of the potential for direct communication via shared environments, we determine the set of program variables that can be used to influence the behaviour of an agent~$\Agent{A}$. We call this set of variables the \emph{influencing variables} of~$\Agent{A}$, which is quite simply the set of all program variables that are referenced in the \levelThree of~$\Agent{A}$ (\ie~$\refV{\Agent{A}}$). Therefore, to directly influence the behaviour of an agent~$\Agent{A}$ via shared environments, an agent must define some variable~$v \in \refV{\Agent{A}}$. 
  
\begin{example}[Computing the Influencing Variables of an Agent]
	Consider the \system described in Section~\ref{sec:example}. The set of influencing variables for the \shipmanager~$\SM{1}$ is given by $\refV{\SM{1}} = \{\var{arriveT[1]}$,~$\var{berthPos[1]}$,~$\var{departT[1]}$, $\var{waitT[1]}\}$. 
This means any agent that defines any of the variables $\var{arriveT[1]}$,~$\var{berthPos[1]}$, $\var{departT[1]}$, or~$\var{waitT[1]}$ will influence the behaviour of~$\SM{1}$.
	The computation of the set of influencing variables for a given system agent can also be computed automatically using our prototype software tool when given the system specification.
\end{example}


\section{Determining Possible Attack Scenarios for Implicit Interactions}
\label{sec:attack_scenarios}
To this point, we have formulated the possible ways in which an agent can directly influence the behaviour of another agent in a given system, either via stimuli or via shared environments. However, implicit interactions are rarely direct interactions, and therefore, we need to generalize the notions of influence that have been established in Sections~\ref{sub:influencing_stimuli} and~\ref{sub:influencing_variables}.

\subsection{Attack Stimuli and Attack Variables}
\label{sub:attack_stimuli_and_variables}

Given an implicit interaction in a system specified using \CCKAabbrv, we want to determine the potential ways in which the interaction can be exploited to mount a cyber-attack. To achieve this, we consider how a compromised source agent can exploit the given implicit interaction via stimuli or via shared environments by examining the pattern of communication (\ie the sequence of direct communications via stimuli and/or shared environments) of the interaction within the given system. As such, we identify the sets of \emph{attack stimuli} and \emph{attack variables} for a given implicit interaction, which are defined by mutual recursion in Definitions~\ref{def:attack_stimuli} and \ref{def:attack_variables}.

\begin{definition}[Attack Stimuli]
\label{def:attack_stimuli}
	Given an implicit interaction of the form~$\impX{n} \deq \impN$, the set of stimuli that a compromised source agent~$\AgentX{n}$ can issue to exploit the implicit interaction and influence the behaviour of the sink agent~$\AgentX{0}$ is given by Equation~\ref{eq:attack-stim}
	%
%
	where~$\attackV{\impX{n-1}}$ denotes the set of attack variables for~$\impX{n-1}$ as defined in Definition~\ref{def:attack_variables}.
\end{definition}

\begin{figure*}[ht]
	\resizebox{0.875\linewidth}{!}{\parbox{\linewidth}{
	\begin{eqnarray}
	\label{eq:attack-stim}
	\attackS{\impX{n}} &=&
	\begin{cases}
		\infl{\AgentX{0}} 	& \hspace{7em}\quad\text{if}\quad \typeX{n} = \stim \nAnd n = 1 \\
		\bigsets{s}{
			\biglnotation{\exists}{a}{a \in \CKAbasic \nAnd a \CKAle\AgentX{n-1} \nAnd s \in \infl{a}}{	  	&\\\hspace{2.5em} 	
				\lnotation{\exists}{v}{}{v \in \defV{\lAct{a}{s}} \STmeet \attackV{\impX{n-1}}}	\Ors		
				\lOut{a}{s} \in \attackS{\impX{n-1}} 
			}}				& \hspace{7em}\quad\text{if}\quad \typeX{n} = \stim \nAnd n > 1 \hspace{2em}\\
		\STbot				& \hspace{7em}\quad\text{otherwise}
	\end{cases}\hspace{4em}\\[1ex]
	\label{eq:attack-var}
	\attackV{\impX{n}} &=&
	\begin{cases}
		\refV{\AgentX{0}} 		& \quad\text{if}\quad \typeX{n} = \env \nAnd n = 1 \\
		\bigsets{v}{
			\biglnotation{\exists}{a}{a \in \CKAbasic \nAnd a \CKAle \AgentX{n-1} \nAnd v \in \refV{a}}{	&\\\hspace{2.5em}
				\lnotation{\exists}{w}{}{w \in \defV{a} \STmeet \attackV{\impX{n-1}}}	\Ors				
				\lnotation{\exists}{s}{s \in \STIMbasic}{\lOut{a}{s} \in \attackS{\impX{n-1}}}
			}}					& \quad\text{if}\quad \typeX{n} = \env \nAnd n > 1 \\
		\STbot					& \quad\text{otherwise}
	\end{cases}
	\end{eqnarray}
	}}
	\vspace{-1.5em}
\end{figure*}

Definition~\ref{def:attack_stimuli} describes the set of stimuli that a compromised source agent can issue to influence the behaviour of the sink agent of a given implicit interaction. To exploit a direct interaction via stimuli~$(\impS{1} \deq \AgentX{1} \toS \AgentX{0})$, a compromised source agent~$\AgentX{1}$ needs to issue any stimulus that influences the sink agent~$\AgentX{0}$. Similarly, to exploit an implicit interaction of the form~$(\impS{n} \deq \AgentX{n} \toS \impX{n-1})$, a compromised source agent~$\AgentX{n}$ needs to issue any atomic stimulus~$s$ for which there exists an atomic sub-behaviour~$a$ of its \emph{neighbouring agent}~$\AgentX{n-1}$ that is influenced by~$s$ and either:
\begin{enumerate}[(a)]
	\item the atomic sub-behaviour~$a$ under~$s$ generates a stimulus that exploits the rest of the given interaction denoted by~$\impX{n-1}$; or
	\item there is a program variable~$v$ that is defined by the resulting behaviour of~$a$ under~$s$ that can exploit the rest of the given interaction denoted by~$\impX{n-1}$. 
\end{enumerate}
Definition~\ref{def:attack_stimuli} also shows that for any implicit interaction of the form~$\AgentX{n} \toE \impX{n-1}$ then~$\attackS{\impX{n}} = \STbot$. This follows from intuition since direct interactions via shared environments (\ie~$\AgentX{n} \toE \AgentX{n-1}$) are exploited only by defining program variables.

\begin{definition}[Attack Variables]
\label{def:attack_variables}
	Given an implicit interaction of the form~$\impX{n} \deq \impN$, the set of variables that a compromised source agent~$\AgentX{n}$ can define to exploit the implicit interaction and influence the behaviour of the sink agent~$\AgentX{0}$ is given by Equation~\ref{eq:attack-var}
%
	where~$\attackS{\impX{n-1}}$ denotes the set of attack stimuli for~$\impX{n-1}$ as defined in Definition~\ref{def:attack_stimuli}.~
\end{definition}

Definition~\ref{def:attack_variables} describes the set of program variables that a compromised source agent can define to influence the behaviour of the sink agent of a given implicit interaction. To exploit a direct interaction via shared environments~$(\impE{1} \deq \AgentX{1} \toE \AgentX{0})$, a compromised source agent~$\AgentX{1}$ needs to define any variable referenced by the sink agent~$\AgentX{0}$. Likewise, to exploit an implicit interaction of the form~$(\impE{n} \deq \AgentX{n} \toE \impX{n-1})$, a compromised source agent~$\AgentX{n}$ needs to define any program variable referenced by an atomic sub-behaviour~$a$ of its \emph{neighbouring agent}~$\AgentX{n-1}$ and for which either:
\begin{enumerate}[(a)]
	\item there is a program variable~$w$ that is defined by the atomic sub-behaviour~$a$ that can exploit the rest of the given interaction denoted by~$\impX{n-1}$; or
	\item there is an atomic stimulus~$s$ for which the atomic sub-behaviour~$a$ under~$s$ generates a stimulus that exploits the rest of the given interaction denoted by~$\impX{n-1}$. 
\end{enumerate}
Similar to Definition~\ref{def:attack_stimuli}, Definition~\ref{def:attack_variables} also shows that for any implicit interaction of the form~$\AgentX{n} \toS \impX{n-1}$ then~$\attackV{\impX{n}} = \STbot$. Once again, this follows from intuition since direct interactions via stimuli (\ie~$\AgentX{n} \toS \AgentX{n-1}$) are exploited only by issuing stimuli.

When determining the attack stimuli and attack variables for the purpose of determining the possible attack scenarios for implicit interactions in Definitions~\ref{def:attack_stimuli} and \ref{def:attack_variables}, we need to consider the existence of an atomic sub-behaviour of the neighbouring agent of~$\AgentX{n}$ (\ie~$a \CKAle \AgentX{n-1}$) to ensure that it is indeed possible for~$\AgentX{n-1}$ to subsequently influence the behaviour of its neighbouring agent. 

As one example of this situation, consider an implicit interaction of the form~$\AgentX{3} \toE \AgentX{2} \toE \AgentX{1} \toE \AgentX{0}$ with~$\agent{A_{2}}{a_1 + a_2}$ such that~$a_1,a_2 \in \CKAbasic$ and the \levelThree{s}~are given by~$a_1 \deq \var{u := v + 1}$ and~$a_2 \deq \var{y := x + 1; w := z}$. Therefore,~$\defV{\AgentX{2}} = \set{\var{u},\var{y},\var{w}}$ and~$\refV{\AgentX{2}} = \set{\var{v},\var{x},\var{z}}$. Now suppose that~$\attackV{\impX{1}} = \set{\var{y},\var{w}}$. By applying Definition~\ref{def:attack_variables}, we compute~$\attackV{\impX{2}} = \set{\var{v},\var{x},\var{z}}$. However, while defining~$\var{v}$ will influence the behaviour of~$\AgentX{2}$ (namely~$a_1$), it will not subsequently influence the behaviour of its neighbouring agent~$\AgentX{1}$ since~$\var{v} \notin \attackV{\impX{1}}$. This means that~$\AgentX{2}$ will only be able to influence the behaviour of~$\AgentX{1}$ if it behaves as the atomic sub-behaviour~$a_2$. Similar cases can be constructed for implicit interactions of different~forms. 


\subsection{Attack Scenario Determination}
\label{sub:attack_scenario_determination}

By combining the definitions of the sets of attack stimuli and attack variables, we obtain a generalized formulation of the set of possible attack scenarios for a given implicit interaction in a distributed system specified using \CCKAabbrv.

\begin{definition}[Attack Scenario Determination]
\label{def:attack_scenario_determinination}
	Given an implicit interaction of the form~$\impX{n} \deq \impN$, the set of possible \emph{attack scenarios} by which a compromised source agent~$\AgentX{n}$ can exploit the implicit interaction and influence the behaviour of the sink agent~$\AgentX{0}$ is given by:
	$\attack{\impX{n}} = \attackS{\impX{n}} \STjoin \attackV{\impX{n}}$.
\end{definition}

By Definition~\ref{def:attack_scenario_determinination}, the set of possible attack scenarios for a given implicit interactions is either the set of attack stimuli or the set of attack variables for the interaction. This is due to the fact that the pattern of communication needs to be respected when exploiting the implicit interaction. This means that the source agent can only influence its immediate neighbour according the type of communication (either via stimuli or shared environments) dictated by the interaction. Consequently, for any given interaction~$\impX{n}$,~$\attack{\impX{n}}$ is either a subset of atomic stimuli or a subset of program variables, depending on the way in which the source agent communicates with its neighbouring agent. This is a direct result from Definitions~\ref{def:attack_stimuli} and~\ref{def:attack_variables}, and the fact that for any implicit interaction~$\impX{n}$: if~$\typeX{n} = \stim$ 
then~$\attackV{\impX{n}} = \STbot$, and if~$\typeX{n} = \env$ then~$\attackS{\impX{n}} = \STbot$.

\pagebreak
\begin{example}[Computing the Possible Attack Scenarios of Implicit Interactions]
	Consider the \system described in Section~\ref{sec:example} and the implicit interaction represented as~$p_7 \deq \SV{1} \toS \SM{2} \toS \PC \toE \SV{2}$. By direct application of Definitions~\ref{def:attack_stimuli}--\ref{def:attack_scenario_determinination}, the set of possible attack scenarios is given by~$\attack{p_7} = \set{\ScompleteB}$. This shows that to exploit the given implicit interaction, a compromised~$\SV{1}$ can send a~$\ScompleteB$ message, which can cause~$\SM{2}$ to enter its leaving behaviour ($\Kleave$) and send a~$\SdepartB$ message. In turn, this can cause~$\PC$ to clear the ship information which is needed by~$\SV{2}$, and can therefore disrupt the port operations.
	
	As another example, consider the implicit interaction represented as~$p_{13} \deq \SV{1} \toE \TM \toE \SV{2}$. In this case, the set of possible attack scenarios is given by~$\attack{p_{13}} = \{\var{berth[1]}$, $\var{berth[2]}$, $\var{numCranes[1]}$, $\var{numCranes[2]}\}$. This shows that a compromised~$\SV{1}$ can exploit the given implicit interaction by modifying any, or all, of the variables~$\var{berth[1]}$,~$\var{berth[2]}$,~$\var{numCranes[1]}$, and/or~$\var{numCranes[2]}$, which are used by~$\TM$ to determine the crane allocations. This means that when~$\TM$ references these variables, it can determine incorrect crane allocations. Therefore, once~$\SV{2}$ enters its planning behaviour ($\Kplan$), it may use the incorrect crane allocations, which can also disrupt the port operations.

Using our prototype software tool, we can automatically compute the set of possible attack scenarios for an implicit interaction when given the system specification. A selection of the results of the tool output are summarized in Table~\ref{tbl:exploitability_results} in Section~\ref{sec:discussion}.
\end{example}

It should be noted that if~$\attack{\impS{n}} = \STbot$ for any implicit interaction of the form~$\impS{n}$, then the implicit interaction can only be exploited trivially. As mentioned in Section~\ref{sub:influencing_stimuli}, every agent other than the inactive agent~$0$ can be influenced by the deactivation stimulus~$\Dstim$. Therefore, if~$\attack{\impS{n}} = \STbot$, then~$\impS{n}$ can only be exploited by issuing the deactivation stimulus~$\Dstim$, which will ultimately cause the sink agent to behave as the inactive agent~$0$. 
Furthermore, if~$\attack{\impE{n}} = \STbot$ for any implicit interaction of the form~$\impE{n}$, then the implicit interaction cannot be exploited. While it is the case that there is a potential for direct communication via shared environments from~$\AgentX{n}$ to~$\AgentX{n-1}$ which allows for the identification of~$\impE{n}$ as an implicit interaction, the attack scenario determination shows that there is no way in which a compromised agent~$\AgentX{n}$ can create a chain of influence to ultimately affect the behaviour of sink agent~$\AgentX{0}$. As an example, consider the implicit interaction represented as~$p_{11} \deq \SV{1} \toE \SM{1} \toS \PC \toS \SM{2} \toE \SV{2}$ for which~$\attack{p_{11}} = \STbot$. By carefully examining the attack scenarios, we find that in order to exploit this implicit interaction,~$\SV{1}$ must define any program variable that is referenced by an atomic sub-behaviour of~$\SM{1}$ for which there is an atomic stimulus that will generate a stimulus that can exploit the rest of the given interaction (\ie~$\SM{1} \toS \PC \toS \SM{2} \toE \SV{2}$), which in this case is the~$\Sarrive$ stimulus. However, because there does not exist any behaviour in the \system that can generate the~$\Sarrive$ stimulus (it is an external stimulus), there does not exist any variable that~$\SV{1}$ can define to cause the chain of influence to ultimately affect the behaviour of~$\SV{2}$ via the given implicit interaction.

Proposition~\ref{prop:attack_failure} shows that if there are no attack scenarios for any suffix of an implicit interaction~$\impX{n}$, then there is no way to exploit~$\impX{n}$ apart from the trivial cases as described above.

\begin{proposition}
\label{prop:attack_failure}
	Let~$\impX{n}$ be an implicit interaction. Then, $\attack{\impX{n-1}} = \STbot \mImp \attack{\impX{n}} = \STbot$ where~$\typeX{i} \in \set{\stim,\env}$ for~$1 < i \le n$.

	\begin{proof}
		The proof follows straightforwardly from Definition~\ref{def:attack_stimuli}, 
		Definition~\ref{def:attack_variables}, and Definition~\ref{def:attack_scenario_determinination}.
		The detailed proof can be found in Appendix~B.
	\end{proof}
\end{proposition}

The result of Proposition~\ref{prop:attack_failure} enables us to determine the attack scenarios for the given implicit interaction without the need to analyze the entire interaction. In cases where a given implicit interaction is long, which is possible in systems with a large numbers of interacting agents, this result allows for savings in terms of the time required to perform the attack scenario determination. 


\section{Evaluating the Exploitability of Implicit Interactions}
\label{sec:exploitability}
After performing the attack scenario determination for each of the identified implicit interactions in a given system design, we use the set of possible attack scenarios to develop a new measure of the severity of each interaction. The severity of an implicit interaction gives an indication of the interactions that have the potential to most negatively impact the safety, security, and/or reliability of the system in which they exist. We call this new measure of severity, the \emph{exploitability} of the implicit interaction and compute it as prescribed by Definition~\ref{def:exploitability}.

\begin{definition}[Exploitability]
\label{def:exploitability}
	The \emph{exploitability} of an implicit interaction~$\impX{n}$ (denoted~$\exploit{\impX{n}}$) is computed recursively by:\\[1ex]
	\resizebox{0.825\linewidth}{!}{\parbox{\linewidth}{
	\begin{equation*}
	\exploit{\impX{n}} =
	\begin{cases}
		\displaystyle\exploit{\impX{n-1}} \frac{\abs{\infl{\AgentX{n-1}} \STmeet \attack{\impX{n}}}}{\abs{\infl{\AgentX{n-1}}}}	
				& \text{if}\quad \typeX{n} = \stim \nAnd n > 1 \\[1em]
		\displaystyle\exploit{\impX{n-1}} \frac{\abs{\refV{\AgentX{n-1}} \STmeet \attack{\impX{n}}}}{\abs{\refV{\AgentX{n-1}}}}	
				& \text{if}\quad \typeX{n} = \env  \nAnd n > 1 \\[1em]
		1		& \text{otherwise}
	\end{cases}
	\end{equation*}
	}}
\end{definition}

Definition~\ref{def:exploitability} computes the fraction of ways that a source agent can influence the behaviour of its neighbouring agent in a way that the influence is propagated along the implicit interaction to eventually influence the behaviour of the sink agent. The exploitability of an implicit interaction interaction~$\impX{n}$ is a numeric value~$\exploit{\impX{n}}$ such that~$0 \le \exploit{\impX{n}} \le 1$. In each fractional component of the exploitability measure, the denominator represents the total number of ways in which the behaviour of the next agent in the interaction can be influenced, and the numerator represents the number of those ways that will maintain the chain of influence for the given the implicit interaction, thereby allowing for its exploitation.

Definition~\ref{def:exploitability} shows that the exploitability of direct interactions is always equal to 1. This follows from intuition, and directly from Definition~\ref{def:attack_scenario_determinination}, since any attack scenario for a direct interaction will influence the behaviour of the sink agent. In the recursive cases, we compute the product of the exploitability of each proper subpath of the given implicit interaction. This allows us to account for the fact that an indirect implicit interaction requires that each intermediate agent propagate the influence to its neighbouring agent. This means that for an implicit interaction that contains a large number of intermediate agents, a compromised source agent needs to rely on a number of additional agents to influence the sink agent's behaviour. Intuitively, the fewer possibilities that each agent in an implicit interaction has to cause a ``chain reaction'' of influence in its neighbouring agents, the lower the exploitability of interaction. In this way, the lower the exploitability measure for an implicit interaction, the more narrow the possibilities for exploiting the interaction.

Note that for an implicit interaction~$\impX{n}$, the exploitability measure is always defined because it is the case that~$\infl{\AgentX{i}} \neq \STbot$ if~$\typeX{i} = \stim$ and~$\refV{\AgentX{i}} \neq \STbot$ if~$\typeX{i} = \env$ for all~$\AgentX{i}$ in~$\impX{n}$ and~$1 \le i \le n$. This follows from the definition of an implicit interaction as a sequence of direct communications either via stimuli or shared environments (see Sections~\ref{sub:influencing_stimuli} and \ref{sub:influencing_variables}). For an implicit interaction to exist, there must be at least one stimulus or program variable that can influence each neighbouring agent in the interaction (\ie a potential for direct communication). For instance, an implicit interaction of the form~$\AgentX{2} \toS \AgentX{1} \toE \AgentX{0}$ is only possible if there exists some stimulus issued by~$\AgentX{2}$ that influences~$\AgentX{1}$ (\ie~$\lnotation{\exists}{s}{}{s \in \infl{\Agent{A_1}}}$), and some program variable defined by~$\AgentX{1}$ that is referenced by~$\AgentX{0}$ (\ie~$\lnotation{\exists}{v}{}{v \in \refV{\Agent{A_0}}}$).

Consider the \system described in Section~\ref{sec:example} and the implicit interaction represented as~$p_7 \deq \SV{1} \toS \SM{2} \toS \PC \toE \SV{2}$. By applying Definition~\ref{def:exploitability}, the exploitability of~$p_7$ is computed to be 0.222. This is due to the fact that, of the three stimuli that will influence the behaviour of~$\SM{2}$, only one (namely~$\ScompleteB$) will allow~$\SM{2}$ to, in turn, influence the behaviour of~$\PC$, and ultimately the rest of the agents in the given interaction. Similarly, of the three stimuli that will influence the behaviour of~$\PC$, only two (namely~$\SdepartA$ and~$\SdepartB$) will allow~$\PC$ to influence the behaviour of~$\SV{2}$.

The exploitability of an implicit interaction can be computed automatically using our prototype software tool when given the system specification. We refer the reader to Table~\ref{tbl:exploitability_results} in Section~\ref{sec:discussion} for a selection of results from the tool output.

\begin{example}[Computing the Exploitability of an Implicit Interaction]
	Consider the \system described in Section~\ref{sec:example} and the implicit interaction represented as~$p_7 \deq \SV{1} \toS \SM{2} \toS \PC \toE \SV{2}$. By applying Definition~\ref{def:exploitability}, the exploitability of~$p_7$ is computed to be 0.222.
	
\resizebox{0.8\linewidth}{!}{\parbox{1.1\linewidth}{
\begin{eqnarray*}
	& & \exploit{\SV{1} \toS \SM{2} \toS \PC \toE \SV{2}} \\
	&=& \exploit{\SM{2} \toS \PC \toE \SV{2}} \quad *	\\
		&&\qquad \frac{\abs{\infl{\SM{2}} \STmeet \attack{\SV{1} \toS \SM{2} \toS \PC \toE \SV{2}}}}{\abs{\infl{\SM{2}}}}	\\
	&=& \exploit{\PC \toE \SV{2}} \quad *	\\
		&&\qquad  \frac{\abs{\infl{\PC} \STmeet \attack{\SM{2} \toS \PC \toE \SV{2}}}}{\abs{\infl{\PC}}} \quad *	\\
		&&\qquad  \frac{\abs{\infl{\SM{2}} \STmeet \attack{\SV{1} \toS \SM{2} \toS \PC \toE \SV{2}}}}{\abs{\infl{\SM{2}}}}	\\
	&=& 1 * \frac{\abs{\set{\Sarrive,\SdepartA,\SdepartB} \STmeet \set{\SdepartA,\SdepartB,\SmanageA,\SmanageB}}}{\abs{\set{\Sarrive,\SdepartA,\SdepartB}}} \quad *	\\
		&&\qquad  \frac{\abs{\set{\Sberth, \ScompleteB, \SmanageB} \STmeet \set{\ScompleteB}}}{\abs{\set{\Sberth, \ScompleteB, \SmanageB}}}	\\
	&=& 1 * \frac{\abs{\set{\SdepartA,\SdepartB}}}{\abs{\set{\Sarrive,\SdepartA,\SdepartB}}}\quad *	\\
		&&\qquad  \frac{\abs{\set{\ScompleteB}}}{\abs{\set{\Sberth,\ScompleteB,\SmanageB}}}\\
	&=& 1 * \frac{2}{3} * \frac{1}{3}\\
	&=& \fbox{0.222}
\end{eqnarray*}}}

This shows that, of the three stimuli that will influence the behaviour of~$\SM{2}$, only one (namely~$\ScompleteB$) will allow~$\SM{2}$ to, in turn, influence the behaviour of~$\PC$, and ultimately the rest of the agents in the given interaction. Similarly, of the three stimuli that will influence the behaviour of~$\PC$, only two (namely~$\SdepartA$ and~$\SdepartB$) will allow~$\PC$ to influence the behaviour of~$\SV{2}$.
The exploitability of an implicit interaction can be computed automatically with the help of our prototype software tool when given the system specification. We refer the reader to Table~\ref{tbl:exploitability_results} in Section~\ref{sec:discussion} for a selection of results from the tool output.
\end{example}

It is important to note that we are determining and measuring the possible ways in which an attacker can use an implicit interaction to influence the behaviour of an agent in a given system, regardless of the impact that such an influence can have on the overall system behaviour. Not all of the ways in which an attacker may exploit an implicit interaction are ``created equal,'' and we do not rule out the fact that some ways may be more likely to be used by an attacker than others, for a number of reasons. Because of this, we acknowledge that the study of the potential impact that particular exploits of existing implicit interactions in system designs can have on the overall system behaviour and operation is critically important, however it is a significant effort in its own right and is out of the scope of this paper. Rather, we conjecture that the information generated from the proposed attack scenario determination and exploitability analysis can provide vital information for studying the potential impact of cyber-attacks launched through implicit interactions and is the subject of our future work.

\section{Experimental Results and Discussion}
\label{sec:discussion}
In this section, we summarize our experimental results for determining the possible attack scenarios and evaluating the exploitability of implicit interactions identified in our illustrative \system. We also provide a discussion of the results and the proposed approach.

\subsection{Experimental Results}
\label{sub:experimental_results}

Using our developed prototype software tool, we compute the attack scenario determination and the exploitability of each of the identified implicit interactions from~$\SV{1}$ to~$\SV{2}$ (see Section~\ref{sub:implicit_interactions} and Fig.~\ref{fig:implicit}). The experimental results are summarized in Table~\ref{tbl:exploitability_results}. A similar analysis can be performed for the remaining implicit interactions identified in the system. Due to space limitations, we do not present the analysis of the entire system here.

\begin{table*}[t]
	\centering
	\caption{Experimental results of the attack scenario determination and exploitability analysis for the identified implicit interactions from~$\SV{1}$ to~$\SV{2}$; a higher exploitability measure indicates a higher threat in the system.}
	\label{tbl:exploitability_results}
	\resizebox{\textwidth}{!}{
	\begin{tabular}{p{0.25in}p{2.25in}p{2.75in}c}
		\textbf{ID} & \textbf{Implicit Interaction} & \textbf{Attack Scenarios:~$\attack{p_i}$}	& \textbf{Exploitability:~$0 \le \exploit{p_i} \le 1$}						  \\\midrule
 		$p_1$		& $\SV{1} \toE \CM \toS \SV{2}$								& $\set{\var{plan},\var{sequence}}$		& 0.333	\\
 		$p_2$		& $\SV{1} \toS \CM \toS \SV{2}$								& $\set{\Supdate}$						& 0.250	\\
 		$p_3$		& $\SV{1} \toS \SM{2} \toS \PC \toE \SM{1} \toS \SV{2}$		& $\set{\ScompleteB}$					& 0.167	\\
 		$p_4$		& $\SV{1} \toS \SM{2} \toS \PC \toS \SM{1} \toS \SV{2}$		& $\STbot$ 								& 0.000	\\
 		$p_5$		& $\SV{1} \toE \SM{1} \toS \PC \toE \SV{2}$					& $\set{\var{berthPos[1]}}$				& 0.167	\\
 		$p_6$		& $\SV{1} \toS \SM{1} \toS \PC \toE \SV{2}$					& $\set{\ScompleteA}$					& 0.222	\\
 		$p_7$		& $\SV{1} \toS \SM{2} \toS \PC \toE \SV{2}$					& $\set{\ScompleteB}$					& 0.222	\\
 		$p_8$		& $\SV{1} \toS \SM{2} \toE \SV{2}$							& $\set{\ScompleteB,\SmanageB}$			& 0.667	\\
 		$p_9$		& $\SV{1} \toE \SM{1} \toS \PC \toE \SM{2} \toE \SV{2}$		& $\set{\var{berthPos[1]}}$				& 0.125	\\
 		$p_{10}$	& $\SV{1} \toS \SM{1} \toS \PC \toE \SM{2} \toE \SV{2}$		& $\set{\ScompleteA}$					& 0.167	\\
 		$p_{11}$	& $\SV{1} \toE \SM{1} \toS \PC \toS \SM{2} \toE \SV{2}$		& $\STbot$ 								& 0.000	\\
 		$p_{12}$	& $\SV{1} \toS \SM{1} \toS \PC \toS \SM{2} \toE \SV{2}$		& $\STbot$ 								& 0.000	\\
 		$p_{13}$	& $\SV{1} \toE \TM \toE \SV{2}$								& $\set{\var{berth[1]},\var{berth[2]},\var{numCranes[1]},\var{numCranes[2]}}$												& 1.000	\\
 		$p_{14}$	& $\SV{1} \toS \TM \toE \SV{2}$								& $\set{\ScompleteA, \ScompleteB, \ScraneA, \ScraneB}$	
																														& 1.000	\\
 		$p_{15}$	& $\SV{1} \toS \SM{2} \toS \SV{2}$							& $\set{\Sberth, \SmanageB}$			& 0.667	\\
 		$p_{16}$	& $\SV{1} \toE \SM{1} \toS \PC \toE \SM{2} \toS \SV{2}$		& $\set{\var{berthPos[1]}}$				& 0.125	\\
 		$p_{17}$	& $\SV{1} \toS \SM{1} \toS \PC \toE \SM{2} \toS \SV{2}$		& $\set{\ScompleteA}$					& 0.167	\\
 		$p_{18}$	& $\SV{1} \toE \SM{1} \toS \PC \toS \SM{2} \toS \SV{2}$		& $\STbot$ 								& 0.000	\\
 		$p_{19}$	& $\SV{1} \toS \SM{1} \toS \PC \toS \SM{2} \toS \SV{2}$		& $\STbot$ 								& 0.000	\\
		\bottomrule
	\end{tabular}}
	\vspace{-1em}
\end{table*}

When comparing the exploitability of the given implicit interactions that have been identified to exist in the \system with a source agent~$\SV{1}$ and a sink agent~$\SV{2}$, we find a variation in the results for each of the interactions. An interaction with a higher exploitability shows that there are more ways in which a compromised source agent can influence the behaviour of the sink agent. This means that such interactions present a higher probability that the source agent can mount a cyber-attack on the given interaction and ultimately influence the behaviour of the sink agent. This analysis aids in validating the existence of the implicit interactions within the system, and provides system designers with plenty of insight into identifying, assessing, and mitigating deficiencies in their designs.

Table~\ref{tbl:exploitability_results} shows that the exploitability measure for some implicit interactions (\eg~$p_{13}$ and~$p_{14}$) is 1.0. This indicates that these implicit interactions are \emph{maximally exploitable}, meaning that as long as a compromised source agent influences the behaviour of its neighbouring agent, then it will ultimately influence the behaviour of the sink agent. As such, this makes the compromised source agent very powerful in being able to conduct a cyber-attack within the system. Consequently, these particular implicit interactions present the most serious threat to the safety, security, and reliability of the system and ought to be assigned the highest priority for mitigation.

Conversely, some implicit interactions (\eg~$p_4$, $p_{11}$, $p_{12}$, $p_{18}$ and~$p_{19}$) have an exploitability measure of 0.0. In the case of~$p_4$,~$p_{12}$, and~$p_{19}$, the results show that these implicit interactions can only be exploited trivially, as discussed in Section~\ref{sub:attack_scenario_determination}, by having the compromised source agent issue a deactivation stimulus~$\Dstim$. In this way, these interactions pose little threat to the system since this very specific and trivial way to exploit the interaction is straightforward to monitor and mitigate. Similarly, in the case of~$p_{11}$ and~$p_{18}$, the results show that these implicit interactions cannot be exploited in the given system. Therefore, these interactions can be considered benign, which is very useful for the system designers when they need to determine where and how to focus their efforts in mitigating the existence of implicit interactions in their system designs.

In addition to the special cases discussed above, Table~\ref{tbl:exploitability_results} also shows that there are a number of implicit interactions with exploitability measures that fall in between the two extremes of not exploitable (\ie~$\exploit{\impE{n}} = 0.0$), or trivially exploitable (\ie~$\exploit{\impS{n}} = 0.0$), and maximally exploitable (\ie~$\exploit{\impX{n}} = 1.0$ for~$\typeX{n} \in \set{\stim,\env}$). For example, the implicit interaction~$p_{8} \deq \SV{1} \toS \SM{2} \toE \SV{2}$ has an exploitability of 0.667. This indicates there is a 66.7\% chance that a compromised~$\SV{1}$ can influence the behaviour of~$\SM{2}$ in such a way that it will ultimately result in an influence of the behaviour of the sink agent~$\SV{2}$. Furthermore, when compared with the implicit interaction~$p_{1} \deq \SV{1} \toE \CM \toS \SV{2}$ which has an exploitability of 0.333 (half of that of~$p_{8}$), we can say that~$p_{8}$ is twice as exploitable as~$p_{1}$. The ability to compare the relative exploitability between two or more implicit interactions can help system designers in determining which implicit interactions found to exist in their designs should be mitigated with the highest priority.

More broadly, the illustrative example of the \system and our experimental results show that despite having two seemingly unconnected components (\eg the \stevedore{s}~$\SV{1}$ and~$\SV{2}$), there is a possibility for one to influence the behaviour of the other. The proposed approach allows us to determine the precise ways in which this is possible with respect to a given system specification. As our experimental results show, in some cases, a compromised source agent requires a very specific scenario to exploit an implicit interaction to influence the behaviour of the sink agent, and in other cases, there is much more freedom and possibility for exploitation. 


\subsection{Discussion of the Proposed Approach}
\label{sub:discussion_of_the_proposed_approach}

The proposed approach for determining the ways in which implicit interactions can be exploited to mount a cyber-attack provides a step towards validating the existence of implicit interactions in the designs of distributed systems. This information is critical in assessing the severity of the vulnerabilities, as well as in determining where and how to spend valuable resources in mitigating the potential for such attacks. In turn, this enables system designers, early in the system development life-cycle, to more accurately assess the threat that such vulnerabilities pose to the overall safety, security, and reliability of the system if left unmitigated. Furthermore, the proposed approach can aid in developing guidelines for designing and implementing resilient distributed systems. For example, it can help designers to rework their system designs to eliminate or mitigate the identified vulnerability and/or to aid in selecting appropriate security and reliability controls (such as strict input validations) to be implemented to prevent any exploits or attacks of vulnerabilities that cannot be completely eliminated.
%
%

Although any sufficiently general model would allow studying interactions between components and their relationship to certain classes of properties, our approach using the \CCKAabbrv modeling framework takes advantage of the capability of \CCKAabbrv to separate the behaviour of a system and its environment, and to deterministically ascertain the potential attack scenarios for both communication via stimuli (message-passing communication) and communication via shared environments (shared variable communication); something that cannot be done directly using other approaches. The moderate effort required to model a given system using \CCKAabbrv (\ie to develop the formal specification of the system) is outweighed by the natural formalization of the notions of the attack scenario determination and  exploitability as presented in this paper. Furthermore, it provides the ability to perform other kinds of analyses (\eg model-checking, simulations, \etc) on the \CCKAabbrv specifications, including those outside of the realm of implicit interactions. 


\section{Concluding Remarks and Future Work}
\label{sec:conclusion}
Implicit interactions are previously unknown linkages among system components indicating the presence of cybersecurity vulnerabilities that, if exploited, can have serious consequences with respect to the safety, security, and reliability of a system. In this paper, we presented a systematic approach for evaluating the exploitability of implicit interactions in distributed systems. The approach is based on \emph{attack scenario determination}, which finds the set of possible ways in which a compromised system agent can exploit a particular implicit interaction to mount a cyber-attack that influences the behaviour of other agents in the system. This is done by studying the influence and response of the system agents and their \CCKAabbrv specifications. We have also developed a new measure of exploitability for implicit interactions, which provides critical information that can offer useful insights to system designers when determining measures to mitigate the potential for implicit interactions to be exploited in a cyber-attack. In addition, we reported on a prototype tool that aids in the automated analysis, and demonstrates the feasibility and practicality of the proposed approach for analyzing systems of reasonable size and complexity. Broadly speaking, the rigorous and practical techniques presented in this paper enable better identification and assessment of cybersecurity vulnerabilities in system designs which can improve overall system resilience, dependability, and security.

While we have shown that there are specific scenarios by which an implicit implicit interaction can be exploited, and that there are varying degrees of exploitability, a further examination and assessment of the impact that a potential cyber-attack can have on a system is needed. For example, while it may be the case that an implicit interaction is highly exploitable, it is possible that the resulting system behaviour from an attack may not lead to a critical system state that is cause for serious concern. As such, in future work, we plan to develop analysis methods based on simulations of cyber-attacks launched upon implicit interactions using the attack scenarios determined by the proposed approach, to study their potential effects and impacts on the given systems and their operations. The results of these simulations and impact analyses will provide actionable information on where to focus efforts and resources on reducing the risk and impact of such attacks.

\section*{Acknowledgment}
This work is supported by the U.S. Department of Homeland Security under Grant Number~2015-ST-061-CIRC01.\\
\textbf{\textrm{Disclaimer}}: The views and conclusions contained in this document are those of the authors and should not be interpreted as necessarily representing the official policies, either expressed or implied, of the U.S. Department of Homeland Security.

\bibliographystyle{ieeetr} 
\bibliography{arXiv2020}

\newpage
\onecolumn
\setcounter{page}{1}
\setcounter{figure}{0}
\setcounter{table}{0}
\section*{Evaluating the Exploitability of Implicit Interactions in Distributed Systems}
\appendices
\section{Algebraic Structures}
\label{app:algebraic_structures}
This appendix summarizes the relevant algebraic structures discussed in the paper.

\begin{enumerate}
	
	\item A \emph{monoid} is a mathematical structure~$\monoid{S}{\cdot}{1}$ where~$S$ is a nonempty set,~$\cdot$ is an associative binary operation and~$1$ is the identity with respect to~$\cdot$ (\ie~$a \cdot 1 = 1 \cdot a = a$ for all~$a \in S$). 
	\begin{itemize}
		\item A monoid is called \emph{commutative} if~$\cdot$ is commutative (\ie~$a \cdot b = b \cdot a$ for all~$a,b \in S$).
		\item A monoid is called \emph{idempotent} if~$\cdot$ is idempotent (\ie~$a \cdot a = a$ for all~$a \in S$).
	\end{itemize}
	
	\item A \emph{semiring} is a mathematical structure~$\semiring{S}{+}{\cdot}{0}{1}$ where~$\monoid{S}{+}{0}$ is a commutative monoid and~$\monoid{S}{\cdot}{1}$ is a monoid such that~$\cdot$ distributes over~$+$ (\ie~$a \cdot (b+c) = a \cdot b + a \cdot c$ and~$(a+b) \cdot c = a \cdot c + b \cdot c$ for all~$a,b,c \in S$).
	\begin{itemize}
		\item Element~$0$ is called \emph{multiplicatively absorbing} if it annihilates~$S$ with respect to~$\cdot$ (\ie~$a \cdot 0 = 0 \cdot a = 0$ for all~$a \in S$). 
		\item A semiring is called \emph{idempotent} if~$+$ is idempotent.
		\item Every idempotent semiring has a partial order~$\le$ on~$S$ defined by~$a \le b \mIff a + b = b$.
	\end{itemize}
	
	\item A \emph{\KA} is a mathematical structure~$\KAstructure$ where~$\KAsemiring$ is an idempotent semiring with a multiplicatively absorbing~$0$ and identity~$1$, and where the following axioms are satisfied for all~$a,b,c \in \CKAset$:
	\begin{enumerate}
	\begin{minipage}[t]{0.45\linewidth}
		\item \label{def:KA_right_unfold}
			$1 + a \cdot \KAstar{a} = \KAstar{a}$
		\item \label{def:KA_left_unfold}
			$1 + \KAstar{a} \cdot a = \KAstar{a}$
	\end{minipage}
	\begin{minipage}[t]{0.55\linewidth}
		\item \label{def:KA_left_induction}
			$b + a \cdot c \le c \mImp \KAstar{a} \cdot b \le c$
		\item \label{def:KA_right_induction}
			$b + c \cdot a \le c \mImp b \cdot \KAstar{a} \le c$
	\end{minipage}
	\end{enumerate}~\\[-1ex]

	\item Let~$\stim = \semiring{\STIMset}{\STIMplus}{\STIMdot}{0_\stim}{1_\stim}$ be a semiring and~$\cka = \monoid{\CKAset}{+}{0_\cka}$ be a commutative monoid. We call~$\Lsemimodule{\stim}{\CKAset}{+}$ a \emph{\leftSemimodule{\stim}} if there exists a mapping~$\lActSig$ such that for all~$s,t \in \STIMset$ and~$a,b \in \CKAset$:
	\begin{enumerate}
	\begin{minipage}[t]{0.45\linewidth}
		\item \label{def:SM_dist_Kplus}
			$\lAct{(a + b)}{s} = \lAct{a}{s} + \lAct{b}{s}$
		\item \label{def:SM_dist_Splus}
			$\lAct{a}{(s \STIMplus t)} = \lAct{a}{s} + \lAct{a}{t}$
		\item \label{def:SM_assoc_seq}
			$\lAct{a}{(s \STIMdot t)} = \lAct{(\lAct{a}{t})}{s}$
	\end{minipage}
	\begin{minipage}[t]{0.55\linewidth}
		\item \label{def:SM_id}
			$\Lsemimodule{\stim}{\CKAset}{\STIMplus}$ is \emph{unitary} if also~$\lAct{a}{1_{\stim}} = a$
		\item \label{def:SM_zero}
			$\Lsemimodule{\stim}{\CKAset}{\STIMplus}$ is \emph{zero-preserving} if also~$\lAct{a}{0_{\stim}} = 0_\cka$ 
	 \end{minipage}
	\end{enumerate}
	\begin{itemize}
		\item An analogous \emph{\rightSemimodule{\cka}} corresponding is denoted by~$\OutSemimodule$. In this paper, we use~$\lOutSig$ to denote the semimodule mapping for~$\OutSemimodule$.
	\end{itemize}

	\item A \emph{\CKA (\CKAabbrv)} is a mathematical structure~$\CKAstructure$ such that~$\CKAstructurePar$ and~$\CKAstructureSeq$ are Kleene algebras linked by the \emph{exchange axiom}~$(a \CKApar b) \CKAseq (c \CKApar d) \le (b \CKAseq c) \CKApar (a \CKAseq d)$.
	\vspace{1ex}
	\begin{itemize}
		\begin{minipage}[t]{0.5\linewidth}
		\item~$\CKAset$ represents a set of possible behaviours.
		\item~$+$ is a choice of two behaviours.
		\item~$\CKAseq$ is a sequential composition of two behaviours.
		\item~$\CKApar$ is a concurrent composition of two behaviours.
		\end{minipage}
		\begin{minipage}[t]{0.5\linewidth}
		\item~$\CKAiterSeq{}$ is a finite sequential iteration of a behaviour.
		\item~$\CKAiterPar{}$ is a finite concurrent iteration of a behaviour.
		\item~$0$ represents the behaviour of the \emph{inactive agent}.
		\item~$1$ represents the behaviour of the \emph{idle agent}.
		\end{minipage}
	\end{itemize}
	
	\item A \emph{stimulus structure}~$\stim \deq \STIMstructure$ is an idempotent semiring with a multiplicatively absorbing~$\Dstim$ and identity~$\Nstim$.
	\begin{itemize}
		\item~$\STIMset$ is the set of stimuli which may be introduced in a system.
		\item~$\STIMplus$ is a choice of two stimuli.
		\item~$\STIMdot$ is a sequential composition of two stimuli.
		\item~$\Dstim$ represents the \emph{deactivation stimulus} which influences all agents to become inactive.
		\item~$\Nstim$ represents the \emph{neutral stimulus} which has no influence on the behaviour of all agents.
	\end{itemize}
	
	\item A \emph{dependence relation} on a set~$\CKAset$ with operator~$+$ is a bilinear relation $\depOp \STleq \CKAset \times \CKAset$ (\ie $\bigB{\dep{c}{(a + b)} \mIff (\dep{c}{a} \Ors \dep{c}{b})}$ and~$\bigB{\dep{(b + c)}{a} \mIff (\dep{b}{a} \Ors \dep{c}{a})}$ for all~$a,b,c \in S$). 
	\begin{itemize}
		\item If~$\dep{b}{a}$, we say that~$b$ \emph{depends on}~$a$.
	\end{itemize}
	
\end{enumerate}

\newpage
\section{Detailed Proofs of Propositions}
\label{app:proofs}
\subsection{Detailed Proof of Proposition~\ref{prop:infl_fixed_point}}
\label{sub:detailed_proof_of_infl_fixed_point}

Let $\agent{A}{a}$ be an agent such that~$a$ is a fixed point behaviour (\ie~$\lnotation{\forall}{s}{s \in \STIMset \STdiff \set{\Dstim}}{\lAct{a}{s} = a}$). Then,~$\infl{\Agent{A}} = \STbot$.

	\Beginproof
		\pred{$\infl{\Agent{A}} = \STbot$}
		$\mIff$ \com{Definition~\ref{def:influencing_stimuli}}
		\pred{$\sets{s \in \STIMbasic}{\lAct{a}{s} \neq a} = \STbot$}
		$\mImpl$ \com{Hypothesis:~$a$ is a fixed point \hsep $\STIMbasic \STleq \STIMset$}
		\pred{$\sets{s \in \STIMbasic}{\false} = \STbot$}
		$\mIff$ \com{Empty Set Axiom \hsep Reflexivity of~$=$}
		\pred{$\true$}
	\Endproof
\vspace{-2em}

\subsection{Detailed Proof of Proposition~\ref{prop:attack_failure}}
\label{sub:detailed_proof_of_attack_failure}

Let~$\impX{n}$ be an implicit interaction. Then,~$\attack{\impX{n-1}} = \STbot \mImp \attack{\impX{n}} = \STbot$ where $\typeX{i} \in \set{\stim,\env}$ for $1 < i \le n$.

	\Beginproof
		\pred{$\attack{\impX{n}} = \STbot$}
		$\mIff$ \com{Definition~\ref{def:attack_scenario_determinination}}
		\pred{$\attackS{\impX{n}} \STjoin \attackV{\impX{n}} = \STbot$}
		$\mIff$ \com{Definition~\ref{def:attack_stimuli} \hsep Definition~\ref{def:attack_variables}}
		\pred{$
		\bigsets{s}{\biglnotation{\exists}{a}{a \in \CKAbasic \nAnd a \CKAle \Agent{X_{n-1}} \nAnd s \in \infl{a}}{	
			\lOut{a}{s} \in \attackS{\impX{n-1}}													\Ors
		 	\lnotation{\exists}{v}{}{v \in \defV{a} \STmeet \attackV{\impX{n-1}}}					}}
		\STjoin 
		\bigsets{v}{\biglnotation{\exists}{a}{a \in \CKAbasic \nAnd a \CKAle \Agent{X_{n-1}} \nAnd v \in \refV{a}}{
			\lnotation{\exists}{w}{}{w \in \defV{a} \STmeet \attackV{\impX{n-1}}}					\Ors
			\lnotation{\exists}{s}{s \in \STIMbasic}{\lOut{a}{s} \in \attackS{\impX{n-1}}}			}}
		 = \STbot$}
		$\mImpl$ \com{Hypothesis: $\attack{\impX{n-1}} = \STbot \mImp \attackS{\impX{n-1}} = \STbot \nAnd \attackV{\impX{n-1}} = \STbot$}
		\pred{$
		\bigsets{s}{\biglnotation{\exists}{a}{a \in \CKAbasic \nAnd a \CKAle \Agent{X_{n-1}} \nAnd s \in \infl{a}}{	
			\lOut{a}{s} \in \STbot														\Ors
		 	\lnotation{\exists}{v}{}{v \in \defV{a} \STmeet \STbot}						}}
		\STjoin 
		\bigsets{v}{\biglnotation{\exists}{a}{a \in \CKAbasic \nAnd a \CKAle \Agent{X_{n-1}} \nAnd v \in \refV{a}}{
			\lnotation{\exists}{w}{}{w \in \defV{a} \STmeet \STbot}					\Ors
			\lnotation{\exists}{s}{s \in \STIMbasic}{\lOut{a}{s} \in \STbot}		}}
		 = \STbot$}
		$\mIff$ \com{Zero of $\STmeet$ \hsep Empty Set Membership}
		\pred{$
		\bigsets{s}{\biglnotation{\exists}{a}{a \in \CKAbasic \nAnd a \CKAle \Agent{X_{n-1}} \nAnd s \in \infl{a}}{	
			\false	\Ors	\lnotation{\exists}{v}{}{\false}	}}
		\STjoin 
		\bigsets{v}{\biglnotation{\exists}{a}{a \in \CKAbasic \nAnd a \CKAle \Agent{X_{n-1}} \nAnd v \in \refV{a}}{
			\lnotation{\exists}{w}{}{\false}	\Ors	\lnotation{\exists}{s}{s \in \STIMbasic}{\false}	}}
		 = \STbot$}
 		$\mIff$ \com{$\exists$-False Body \hsep Identity of $\Ors$}
 		\pred{$\bigsets{s}{\false} \STjoin \bigsets{v}{\false} = \STbot$}
 		$\mIff$ \com{Empty Set Axiom \hsep Reflexivity of~$=$}
		\pred{$\true$}
	\Endproof
\vspace{-2em}

\newpage
\section{\CCKAabbrv Specification of the \SYSTEM}
\label{app:specification}
This appendix contains the complete specification of the \system described in Section~\ref{sec:example}.

\subsection{System Agents}
\label{sub:system_agents}

The \system consists of the following eight agents:\\

\begin{tabular}{clcclccl}
	$\PC$		& \portcaptain 			&\quad\qquad&
	$\SM{1}$ 	& \shipmanager 1		&\quad\qquad&
	$\SM{2}$ 	& \shipmanager 2		\\
	$\TM~$		& \terminalmanager		&\quad\qquad&
	$\SV{1}$ 	& \stevedore 1			&\quad\qquad&
	$\SV{2}$ 	& \stevedore 2			\\
	$\CM~$		& \crane				&\quad\qquad&
	$\CC~$		& \carrier				&\quad\qquad& 
\end{tabular}


\subsection{Stimulus Structure}
\label{sub:stimulus_structure}

The set of stimuli~$\STIMset$ is generated using the operations of stimulus structures and the following set of 21 atomic stimuli:~\PCTstimuli.


\subsection{Behavior (\CKAabbrv) Structure}
\label{sub:behaviour_structure}

The set of agent behaviours~$\CKAset$ is generated using the operations of \CKAabbrv and the following set of 25 atomic behaviours:~\PCTbehaviors.


\subsection{\levelONE{s} of Agents}
\label{sub:stimulus_response_specifications}

The \levelOne{s} for the system agents are provided in a tabular format as shown in Tables~\ref{tbl:agent_PC}--\ref{tbl:agent_CC}. While the \levelOne{s} of the system agents specifies a single next behaviour mapping~$\actOp$ and next stimulus mapping~$\outOp$, they are presented as separate tables for each agent to improve the readability and reviewability of the specifications. For each table, the row header shows the atomic behaviours that the given agent can exhibit as dictated by the \levelTwo of the agent and the \CCKAabbrv. The column header shows the atomic stimuli to which the agent may be subjected. The table grid provides the resulting next behaviour or next stimulus (with respect to the operator given in the top left cell) when the stimulus shown in the column header is applied to the behaviour shown in the row header.

\subsection{\levelTWO{s}}
\label{sub:abstract_behaviour_specifications}

The \levelTwo{s} for the system agents are given in Fig.~\ref{fig:abstract_behaviour}.
The behaviour of the \system ($\Agent{PCT}$) can be represented by the concurrent composition of the behaviours of each of the system agents, \ie 
\begin{equation*}
	\Agent{PCT} \mapsto 
		\bigA{
			\PC 	\CKApar
			\SM{1} 	\CKApar
			\SM{2} 	\CKApar
			\SV{1} 	\CKApar
			\SV{2} 	\CKApar
			\TM 	\CKApar
			\CM 	\CKApar
			\CC 				
		} 
\end{equation*}

\begin{figure}[ht!]
	\centering
	\begin{eqnarray*}
		\PC 	&\mapsto& \bigA{(\KmanA + \KmanB) \CKAseq \Kinit + (\KclearA + \KclearB) \CKAseq \Kdepart} 	\\
		\SM{i} 	&\mapsto& \bigA{\KtimeS + \Kpos + \Kleave}												\\
		\SV{i} 	&\mapsto& \bigA{\Kcrane + \Kplan + \Kdock + \Krlse}									\\
		\TM 	&\mapsto& \bigA{\Kallo + \Kfree}															\\
		\CM 	&\mapsto& \bigA{\Kread \CKAseq \Kcargo + \Kseq \CKAseq \Kserve + \Kupdate \CKAseq \Koperate}	\\
		\CC 	&\mapsto& \bigA{\Kavail \CKAseq \Kassign + \Knear \CKAseq \Kmove}
	\end{eqnarray*}
	\caption{Abstract behaviour specification of the \system agents.}
	\label{fig:abstract_behaviour}
\end{figure}


\pagebreak
\begin{landscape}
\begin{table}[ht!]
	\caption{Stimulus-response specification of the \portcaptain~$\PC$.}
	\label{tbl:agent_PC}
	\resizebox{1.35\textwidth}{!}{
	\begin{tabular}{|p{1.15cm}||p{1.15cm}|p{1.15cm}|p{1.15cm}|p{1.15cm}|p{1.15cm}|p{1.15cm}|p{1.15cm}|p{1.15cm}|p{1.15cm}|p{1.15cm}
					|p{1.15cm}| p{1.15cm}|p{1.15cm}|p{1.15cm}|p{1.15cm}|p{1.15cm}|p{1.15cm}|p{1.15cm}|p{1.15cm}|p{1.15cm}|p{1.15cm}|} \hline
	$\actOp$ 
		&~$\Sarrive$ &~$\SmanageA$ &~$\SmanageB$ &~$\SshipA$ &~$\SshipB$ &~$\ScraneA$ &~$\ScraneB$ &~$\Sallocate$ &~$\Sberth$ &~$\Sdock$ 
		&~$\SoperateA$ &~$\SoperateB$ &~$\Scarrier$ &~$\Sassign$ &~$\Sserve$ &~$\Supdate$ &~$\Sdone$ &~$\ScompleteA$ &~$\ScompleteB$ 
		&~$\SdepartA$ &~$\SdepartB$ \\ \hline\hline
	
	$\KmanA$ 
		&~$\KmanA$ &~$\KmanA$ &~$\KmanA$ &~$\KmanA$ &~$\KmanA$ &~$\KmanA$ &~$\KmanA$ &~$\KmanA$ &~$\KmanA$ &~$\KmanA$ &~$\KmanA$ 
		&~$\KmanA$ &~$\KmanA$ &~$\KmanA$ &~$\KmanA$ &~$\KmanA$ &~$\KmanA$ &~$\KmanA$ &~$\KmanA$ &~$\KclearA$ &~$\KmanA$	\\ \hline
		
	$\KmanB$ 
		&~$\KmanB$ &~$\KmanB$ &~$\KmanB$ &~$\KmanB$ &~$\KmanB$ &~$\KmanB$ &~$\KmanB$ &~$\KmanB$ &~$\KmanB$ &~$\KmanB$ &~$\KmanB$ 
		&~$\KmanB$ &~$\KmanB$ &~$\KmanB$ &~$\KmanB$ &~$\KmanB$ &~$\KmanB$ &~$\KmanB$ &~$\KmanB$ &~$\KmanB$ &~$\KclearB$	\\ \hline
		
	$\Kinit$ 
		&~$\Kinit$ &~$\Kinit$ &~$\Kinit$ &~$\Kinit$ &~$\Kinit$ &~$\Kinit$ &~$\Kinit$ &~$\Kinit$ &~$\Kinit$ &~$\Kinit$ &~$\Kinit$ 
		&~$\Kinit$ &~$\Kinit$ &~$\Kinit$ &~$\Kinit$ &~$\Kinit$ &~$\Kinit$ &~$\Kinit$ &~$\Kinit$ &~$\Kdepart$ &~$\Kdepart$	\\ \hline
		
	$\KclearA$ 
		&~$\KmanA$ &~$\KclearA$ &~$\KclearA$ &~$\KclearA$ &~$\KclearA$ &~$\KclearA$ &~$\KclearA$ &~$\KclearA$ &~$\KclearA$ &~$\KclearA$ &~$\KclearA$ 
		&~$\KclearA$ &~$\KclearA$ &~$\KclearA$ &~$\KclearA$ &~$\KclearA$ &~$\KclearA$ &~$\KclearA$ &~$\KclearA$ &~$\KclearA$ &~$\KclearA$	\\ \hline

	$\KclearB$ 
		&~$\KmanB$ &~$\KclearB$ &~$\KclearB$ &~$\KclearB$ &~$\KclearB$ &~$\KclearB$ &~$\KclearB$ &~$\KclearB$ &~$\KclearB$ &~$\KclearB$ &~$\KclearB$ 
		&~$\KclearB$ &~$\KclearB$ &~$\KclearB$ &~$\KclearB$ &~$\KclearB$ &~$\KclearB$ &~$\KclearB$ &~$\KclearB$ &~$\KclearB$ &~$\KclearB$	\\ \hline
		
	$\Kdepart$ 
		&~$\Kdepart$ &~$\Kinit$ &~$\Kinit$ &~$\Kdepart$ &~$\Kdepart$ &~$\Kdepart$ &~$\Kdepart$ &~$\Kdepart$ &~$\Kdepart$ &~$\Kdepart$ &~$\Kdepart$ 
		&~$\Kdepart$ &~$\Kdepart$ &~$\Kdepart$ &~$\Kdepart$ &~$\Kdepart$ &~$\Kdepart$ &~$\Kdepart$ &~$\Kdepart$ &~$\Kdepart$ &~$\Kdepart$	\\ \hline

\multicolumn{2}{c}{} \\\hline

	$\outOp$ 
		&~$\Sarrive$ &~$\SmanageA$ &~$\SmanageB$ &~$\SshipA$ &~$\SshipB$ &~$\ScraneA$ &~$\ScraneB$ &~$\Sallocate$ &~$\Sberth$ &~$\Sdock$ 
		&~$\SoperateA$ &~$\SoperateB$ &~$\Scarrier$ &~$\Sassign$ &~$\Sserve$ &~$\Supdate$ &~$\Sdone$ &~$\ScompleteA$ &~$\ScompleteB$ 
		&~$\SdepartA$ &~$\SdepartB$ \\ \hline\hline
		
	$\KmanA$ 
		&~$\Nstim$ &~$\Nstim$ &~$\Nstim$ &~$\Nstim$ &~$\Nstim$ &~$\Nstim$ &~$\Nstim$ &~$\Nstim$ &~$\Nstim$ &~$\Nstim$ &~$\Nstim$ 
		&~$\Nstim$ &~$\Nstim$ &~$\Nstim$ &~$\Nstim$ &~$\Nstim$ &~$\Nstim$ &~$\Nstim$ &~$\Nstim$ &~$\SdepartA$ &~$\Nstim$	\\ \hline
		
	$\KmanB$ 
		&~$\Nstim$ &~$\Nstim$ &~$\Nstim$ &~$\Nstim$ &~$\Nstim$ &~$\Nstim$ &~$\Nstim$ &~$\Nstim$ &~$\Nstim$ &~$\Nstim$ &~$\Nstim$ 
		&~$\Nstim$ &~$\Nstim$ &~$\Nstim$ &~$\Nstim$ &~$\Nstim$ &~$\Nstim$ &~$\Nstim$ &~$\Nstim$ &~$\Nstim$ &~$\SdepartB$	\\ \hline

	$\Kinit$ 
		&~$\Nstim$ &~$\Nstim$ &~$\Nstim$ &~$\Nstim$ &~$\Nstim$ &~$\Nstim$ &~$\Nstim$ &~$\Nstim$ &~$\Nstim$ &~$\Nstim$ &~$\Nstim$ 
		&~$\Nstim$ &~$\Nstim$ &~$\Nstim$ &~$\Nstim$ &~$\Nstim$ &~$\Nstim$ &~$\Nstim$ &~$\Nstim$ &~$\Nstim$ &~$\Nstim$	\\ \hline
		
	$\KclearA$ 
		&~$\SmanageA$ &~$\Nstim$ &~$\Nstim$ &~$\Nstim$ &~$\Nstim$ &~$\Nstim$ &~$\Nstim$ &~$\Nstim$ &~$\Nstim$ &~$\Nstim$ &~$\Nstim$ 
		&~$\Nstim$ &~$\Nstim$ &~$\Nstim$ &~$\Nstim$ &~$\Nstim$ &~$\Nstim$ &~$\Nstim$ &~$\Nstim$ &~$\Nstim$ &~$\Nstim$	\\ \hline

	$\KclearB$ 
		&~$\SmanageB$ &~$\Nstim$ &~$\Nstim$ &~$\Nstim$ &~$\Nstim$ &~$\Nstim$ &~$\Nstim$ &~$\Nstim$ &~$\Nstim$ &~$\Nstim$ &~$\Nstim$ 
		&~$\Nstim$ &~$\Nstim$ &~$\Nstim$ &~$\Nstim$ &~$\Nstim$ &~$\Nstim$ &~$\Nstim$ &~$\Nstim$ &~$\Nstim$ &~$\Nstim$	\\ \hline
		
	$\Kdepart$ 
		&~$\Nstim$ &~$\SmanageA$ &~$\SmanageB$ &~$\Nstim$ &~$\Nstim$ &~$\Nstim$ &~$\Nstim$ &~$\Nstim$ &~$\Nstim$ &~$\Nstim$ &~$\Nstim$ 
		&~$\Nstim$ &~$\Nstim$ &~$\Nstim$ &~$\Nstim$ &~$\Nstim$ &~$\Nstim$ &~$\Nstim$ &~$\Nstim$ &~$\Nstim$ &~$\Nstim$	\\ \hline

	 \end{tabular}}
\end{table}
\begin{table}[ht!]
	\caption{Stimulus-response specification of the \shipmanager~$\SM{1}$.}
	\label{tbl:agent_SMa}
	\resizebox{1.35\textwidth}{!}{
	\begin{tabular}{|p{1.15cm}||p{1.15cm}|p{1.15cm}|p{1.15cm}|p{1.15cm}|p{1.15cm}|p{1.15cm}|p{1.15cm}|p{1.15cm}|p{1.15cm}|p{1.15cm}
					|p{1.15cm}| p{1.15cm}|p{1.15cm}|p{1.15cm}|p{1.15cm}|p{1.15cm}|p{1.15cm}|p{1.15cm}|p{1.15cm}|p{1.15cm}|p{1.15cm}|} \hline
	$\actOp$ 
		&~$\Sarrive$ &~$\SmanageA$ &~$\SmanageB$ &~$\SshipA$ &~$\SshipB$ &~$\ScraneA$ &~$\ScraneB$ &~$\Sallocate$ &~$\Sberth$ &~$\Sdock$ 
		&~$\SoperateA$ &~$\SoperateB$ &~$\Scarrier$ &~$\Sassign$ &~$\Sserve$ &~$\Supdate$ &~$\Sdone$ &~$\ScompleteA$ &~$\ScompleteB$ 
		&~$\SdepartA$ &~$\SdepartB$ \\ \hline\hline
	
	$\KtimeS$ 
		&~$\KtimeS$ &~$\KtimeS$ &~$\KtimeS$ &~$\KtimeS$ &~$\KtimeS$ &~$\KtimeS$ &~$\KtimeS$ &~$\KtimeS$ &~$\Kpos$ &~$\KtimeS$ &~$\KtimeS$ 
		&~$\KtimeS$ &~$\KtimeS$ &~$\KtimeS$ &~$\KtimeS$ &~$\KtimeS$ &~$\KtimeS$ &~$\KtimeS$ &~$\KtimeS$ &~$\KtimeS$ &~$\KtimeS$	\\ \hline

	$\Kpos$ 
		&~$\Kpos$ &~$\Kpos$ &~$\Kpos$ &~$\Kpos$ &~$\Kpos$ &~$\Kpos$ &~$\Kpos$ &~$\Kpos$ &~$\Kpos$ &~$\Kpos$ &~$\Kpos$ 
		&~$\Kpos$ &~$\Kpos$ &~$\Kpos$ &~$\Kpos$ &~$\Kpos$ &~$\Kpos$ &~$\Kleave$ &~$\Kpos$ &~$\Kpos$ &~$\Kpos$	\\ \hline

	$\Kleave$ 
		&~$\Kleave$ &~$\KtimeS$ &~$\Kleave$ &~$\Kleave$ &~$\Kleave$ &~$\Kleave$ &~$\Kleave$ &~$\Kleave$ &~$\Kleave$ &~$\Kleave$ &~$\Kleave$ 
		&~$\Kleave$ &~$\Kleave$ &~$\Kleave$ &~$\Kleave$ &~$\Kleave$ &~$\Kleave$ &~$\Kleave$ &~$\Kleave$ &~$\Kleave$ &~$\Kleave$	\\ \hline

\multicolumn{2}{c}{} \\\hline

	$\outOp$ 
		&~$\Sarrive$ &~$\SmanageA$ &~$\SmanageB$ &~$\SshipA$ &~$\SshipB$ &~$\ScraneA$ &~$\ScraneB$ &~$\Sallocate$ &~$\Sberth$ &~$\Sdock$ 
		&~$\SoperateA$ &~$\SoperateB$ &~$\Scarrier$ &~$\Sassign$ &~$\Sserve$ &~$\Supdate$ &~$\Sdone$ &~$\ScompleteA$ &~$\ScompleteB$ 
		&~$\SdepartA$ &~$\SdepartB$ \\ \hline\hline
		
	$\KtimeS$ 
		&~$\Nstim$ &~$\Nstim$ &~$\Nstim$ &~$\Nstim$ &~$\Nstim$ &~$\Nstim$ &~$\Nstim$ &~$\Nstim$ &~$\Sdock$ &~$\Nstim$ &~$\Nstim$ 
		&~$\Nstim$ &~$\Nstim$ &~$\Nstim$ &~$\Nstim$ &~$\Nstim$ &~$\Nstim$ &~$\Nstim$ &~$\Nstim$ &~$\Nstim$ &~$\Nstim$	\\ \hline

	$\Kpos$ 
		&~$\Nstim$ &~$\Nstim$ &~$\Nstim$ &~$\Nstim$ &~$\Nstim$ &~$\Nstim$ &~$\Nstim$ &~$\Nstim$ &~$\Nstim$ &~$\Nstim$ &~$\Nstim$ 
		&~$\Nstim$ &~$\Nstim$ &~$\Nstim$ &~$\Nstim$ &~$\Nstim$ &~$\Nstim$ &~$\SdepartA$ &~$\Nstim$ &~$\Nstim$ &~$\Nstim$	\\ \hline

	$\Kleave$ 
		&~$\Nstim$ &~$\SshipA$ &~$\Nstim$ &~$\Nstim$ &~$\Nstim$ &~$\Nstim$ &~$\Nstim$ &~$\Nstim$ &~$\Nstim$ &~$\Nstim$ &~$\Nstim$ 
		&~$\Nstim$ &~$\Nstim$ &~$\Nstim$ &~$\Nstim$ &~$\Nstim$ &~$\Nstim$ &~$\Nstim$ &~$\Nstim$ &~$\Nstim$ &~$\Nstim$	\\ \hline

	 \end{tabular}}
\end{table}
\begin{table}[ht!]
	\caption{Stimulus-response specification of the \shipmanager~$\SM{2}$.}
	\label{tbl:agent_SMb}
	\resizebox{1.35\textwidth}{!}{
	\begin{tabular}{|p{1.15cm}||p{1.15cm}|p{1.15cm}|p{1.15cm}|p{1.15cm}|p{1.15cm}|p{1.15cm}|p{1.15cm}|p{1.15cm}|p{1.15cm}|p{1.15cm}
					|p{1.15cm}| p{1.15cm}|p{1.15cm}|p{1.15cm}|p{1.15cm}|p{1.15cm}|p{1.15cm}|p{1.15cm}|p{1.15cm}|p{1.15cm}|p{1.15cm}|} \hline
	$\actOp$ 
		&~$\Sarrive$ &~$\SmanageA$ &~$\SmanageB$ &~$\SshipA$ &~$\SshipB$ &~$\ScraneA$ &~$\ScraneB$ &~$\Sallocate$ &~$\Sberth$ &~$\Sdock$ 
		&~$\SoperateA$ &~$\SoperateB$ &~$\Scarrier$ &~$\Sassign$ &~$\Sserve$ &~$\Supdate$ &~$\Sdone$ &~$\ScompleteA$ &~$\ScompleteB$ 
		&~$\SdepartA$ &~$\SdepartB$ \\ \hline\hline
	
	$\KtimeS$ 
		&~$\KtimeS$ &~$\KtimeS$ &~$\KtimeS$ &~$\KtimeS$ &~$\KtimeS$ &~$\KtimeS$ &~$\KtimeS$ &~$\KtimeS$ &~$\Kpos$ &~$\KtimeS$ &~$\KtimeS$ 
		&~$\KtimeS$ &~$\KtimeS$ &~$\KtimeS$ &~$\KtimeS$ &~$\KtimeS$ &~$\KtimeS$ &~$\KtimeS$ &~$\KtimeS$ &~$\KtimeS$ &~$\KtimeS$	\\ \hline

	$\Kpos$ 
		&~$\Kpos$ &~$\Kpos$ &~$\Kpos$ &~$\Kpos$ &~$\Kpos$ &~$\Kpos$ &~$\Kpos$ &~$\Kpos$ &~$\Kpos$ &~$\Kpos$ &~$\Kpos$ 
		&~$\Kpos$ &~$\Kpos$ &~$\Kpos$ &~$\Kpos$ &~$\Kpos$ &~$\Kpos$ &~$\Kpos$ &~$\Kleave$ &~$\Kpos$ &~$\Kpos$	\\ \hline

	$\Kleave$ 
		&~$\Kleave$ &~$\Kleave$ &~$\KtimeS$ &~$\Kleave$ &~$\Kleave$ &~$\Kleave$ &~$\Kleave$ &~$\Kleave$ &~$\Kleave$ &~$\Kleave$ &~$\Kleave$ 
		&~$\Kleave$ &~$\Kleave$ &~$\Kleave$ &~$\Kleave$ &~$\Kleave$ &~$\Kleave$ &~$\Kleave$ &~$\Kleave$ &~$\Kleave$ &~$\Kleave$	\\ \hline

\multicolumn{2}{c}{} \\\hline

	$\outOp$ 
		&~$\Sarrive$ &~$\SmanageA$ &~$\SmanageB$ &~$\SshipA$ &~$\SshipB$ &~$\ScraneA$ &~$\ScraneB$ &~$\Sallocate$ &~$\Sberth$ &~$\Sdock$ 
		&~$\SoperateA$ &~$\SoperateB$ &~$\Scarrier$ &~$\Sassign$ &~$\Sserve$ &~$\Supdate$ &~$\Sdone$ &~$\ScompleteA$ &~$\ScompleteB$ 
		&~$\SdepartA$ &~$\SdepartB$ \\ \hline\hline
		
	$\KtimeS$ 
		&~$\Nstim$ &~$\Nstim$ &~$\Nstim$ &~$\Nstim$ &~$\Nstim$ &~$\Nstim$ &~$\Nstim$ &~$\Nstim$ &~$\Sdock$ &~$\Nstim$ &~$\Nstim$ 
		&~$\Nstim$ &~$\Nstim$ &~$\Nstim$ &~$\Nstim$ &~$\Nstim$ &~$\Nstim$ &~$\Nstim$ &~$\Nstim$ &~$\Nstim$ &~$\Nstim$	\\ \hline

	$\Kpos$ 
		&~$\Nstim$ &~$\Nstim$ &~$\Nstim$ &~$\Nstim$ &~$\Nstim$ &~$\Nstim$ &~$\Nstim$ &~$\Nstim$ &~$\Nstim$ &~$\Nstim$ &~$\Nstim$ 
		&~$\Nstim$ &~$\Nstim$ &~$\Nstim$ &~$\Nstim$ &~$\Nstim$ &~$\Nstim$ &~$\Nstim$ &~$\SdepartB$ &~$\Nstim$ &~$\Nstim$	\\ \hline

	$\Kleave$ 
		&~$\Nstim$ &~$\Nstim$ &~$\SshipB$ &~$\Nstim$ &~$\Nstim$ &~$\Nstim$ &~$\Nstim$ &~$\Nstim$ &~$\Nstim$ &~$\Nstim$ &~$\Nstim$ 
		&~$\Nstim$ &~$\Nstim$ &~$\Nstim$ &~$\Nstim$ &~$\Nstim$ &~$\Nstim$ &~$\Nstim$ &~$\Nstim$ &~$\Nstim$ &~$\Nstim$	\\ \hline

	 \end{tabular}}
\end{table}
\begin{table}[ht!]
	\caption{Stimulus-response specification of the \stevedore~$\SV{1}$.}
	\label{tbl:agent_SVa}
	\resizebox{1.35\textwidth}{!}{
	\begin{tabular}{|p{1.15cm}||p{1.15cm}|p{1.15cm}|p{1.15cm}|p{1.15cm}|p{1.15cm}|p{1.15cm}|p{1.15cm}|p{1.15cm}|p{1.15cm}|p{1.15cm}
					|p{1.15cm}| p{1.15cm}|p{1.15cm}|p{1.15cm}|p{1.15cm}|p{1.15cm}|p{1.15cm}|p{1.15cm}|p{1.15cm}|p{1.15cm}|p{1.15cm}|} \hline
	$\actOp$ 
		&~$\Sarrive$ &~$\SmanageA$ &~$\SmanageB$ &~$\SshipA$ &~$\SshipB$ &~$\ScraneA$ &~$\ScraneB$ &~$\Sallocate$ &~$\Sberth$ &~$\Sdock$ 
		&~$\SoperateA$ &~$\SoperateB$ &~$\Scarrier$ &~$\Sassign$ &~$\Sserve$ &~$\Supdate$ &~$\Sdone$ &~$\ScompleteA$ &~$\ScompleteB$ 
		&~$\SdepartA$ &~$\SdepartB$ \\ \hline\hline
	
	$\Kcrane$ 
		&~$\Kcrane$ &~$\Kcrane$ &~$\Kcrane$ &~$\Kcrane$ &~$\Kcrane$ &~$\Kcrane$ &~$\Kcrane$ &~$\Kplan$ &~$\Kcrane$ &~$\Kcrane$ &~$\Kcrane$ 
		&~$\Kcrane$ &~$\Kcrane$ &~$\Kcrane$ &~$\Kcrane$ &~$\Kcrane$ &~$\Kcrane$ &~$\Kcrane$ &~$\Kcrane$ &~$\Kcrane$ &~$\Kcrane$	\\ \hline	
	
	$\Kplan$ 
		&~$\Kplan$ &~$\Kplan$ &~$\Kplan$ &~$\Kcrane$ &~$\Kplan$ &~$\Kplan$ &~$\Kplan$ &~$\Kplan$ &~$\Kplan$ &~$\Kdock$ &~$\Kplan$ 
		&~$\Kplan$ &~$\Kplan$ &~$\Kplan$ &~$\Kplan$ &~$\Kplan$ &~$\Kplan$ &~$\Kplan$ &~$\Kplan$ &~$\Kplan$ &~$\Kplan$	\\ \hline

	$\Kdock$ 
		&~$\Kdock$ &~$\Kdock$ &~$\Kdock$ &~$\Kcrane$ &~$\Kdock$ &~$\Kdock$ &~$\Kdock$ &~$\Kdock$ &~$\Kdock$ &~$\Kdock$ &~$\Kdock$ 
		&~$\Kdock$ &~$\Kdock$ &~$\Kdock$ &~$\Kdock$ &~$\Kdock$ &~$\Krlse$ &~$\Kdock$ &~$\Kdock$ &~$\Kdock$ &~$\Kdock$	\\ \hline

	$\Krlse$ 
		&~$\Krlse$ &~$\Krlse$ &~$\Krlse$ &~$\Kcrane$ &~$\Krlse$ &~$\Krlse$ &~$\Krlse$ &~$\Krlse$ &~$\Krlse$ &~$\Krlse$ &~$\Krlse$ 
		&~$\Krlse$ &~$\Krlse$ &~$\Krlse$ &~$\Krlse$ &~$\Krlse$ &~$\Krlse$ &~$\Krlse$ &~$\Krlse$ &~$\Krlse$ &~$\Krlse$	\\ \hline

\multicolumn{2}{c}{} \\\hline

	$\outOp$ 
		&~$\Sarrive$ &~$\SmanageA$ &~$\SmanageB$ &~$\SshipA$ &~$\SshipB$ &~$\ScraneA$ &~$\ScraneB$ &~$\Sallocate$ &~$\Sberth$ &~$\Sdock$ 
		&~$\SoperateA$ &~$\SoperateB$ &~$\Scarrier$ &~$\Sassign$ &~$\Sserve$ &~$\Supdate$ &~$\Sdone$ &~$\ScompleteA$ &~$\ScompleteB$ 
		&~$\SdepartA$ &~$\SdepartB$ \\ \hline\hline
		
	$\Kcrane$ 
		&~$\Nstim$ &~$\Nstim$ &~$\Nstim$ &~$\Nstim$ &~$\Nstim$ &~$\Nstim$ &~$\Nstim$ &~$\Sberth$ &~$\Nstim$ &~$\Nstim$ &~$\Nstim$ 
		&~$\Nstim$ &~$\Nstim$ &~$\Nstim$ &~$\Nstim$ &~$\Nstim$ &~$\Nstim$ &~$\Nstim$ &~$\Nstim$ &~$\Nstim$ &~$\Nstim$	\\ \hline	
	
	$\Kplan$ 
		&~$\Nstim$ &~$\Nstim$ &~$\Nstim$ &~$\ScraneA$ &~$\Nstim$ &~$\Nstim$ &~$\Nstim$ &~$\Nstim$ &~$\Nstim$ &~$\SoperateA$ &~$\Nstim$ 
		&~$\Nstim$ &~$\Nstim$ &~$\Nstim$ &~$\Nstim$ &~$\Nstim$ &~$\Nstim$ &~$\Nstim$ &~$\Nstim$ &~$\Nstim$ &~$\Nstim$	\\ \hline

	$\Kdock$ 
		&~$\Nstim$ &~$\Nstim$ &~$\Nstim$ &~$\ScraneA$ &~$\Nstim$ &~$\Nstim$ &~$\Nstim$ &~$\Nstim$ &~$\Nstim$ &~$\Nstim$ &~$\Nstim$ 
		&~$\Nstim$ &~$\Nstim$ &~$\Nstim$ &~$\Nstim$ &~$\Nstim$ &~$\ScompleteA$ &~$\Nstim$ &~$\Nstim$ &~$\Nstim$ &~$\Nstim$	\\ \hline
	
	$\Krlse$ 
		&~$\Nstim$ &~$\Nstim$ &~$\Nstim$ &~$\ScraneA$ &~$\Nstim$ &~$\Nstim$ &~$\Nstim$ &~$\Nstim$ &~$\Nstim$ &~$\Nstim$ &~$\Nstim$ 
		&~$\Nstim$ &~$\Nstim$ &~$\Nstim$ &~$\Nstim$ &~$\Nstim$ &~$\Nstim$ &~$\Nstim$ &~$\Nstim$ &~$\Nstim$ &~$\Nstim$	\\ \hline		

	 \end{tabular}}
\end{table}
\begin{table}[ht!]
	\caption{Stimulus-response specification of the \stevedore~$\SV{2}$.}
	\label{tbl:agent_SVb}
	\resizebox{1.35\textwidth}{!}{
	\begin{tabular}{|p{1.15cm}||p{1.15cm}|p{1.15cm}|p{1.15cm}|p{1.15cm}|p{1.15cm}|p{1.15cm}|p{1.15cm}|p{1.15cm}|p{1.15cm}|p{1.15cm}
					|p{1.15cm}| p{1.15cm}|p{1.15cm}|p{1.15cm}|p{1.15cm}|p{1.15cm}|p{1.15cm}|p{1.15cm}|p{1.15cm}|p{1.15cm}|p{1.15cm}|} \hline
	$\actOp$ 
		&~$\Sarrive$ &~$\SmanageA$ &~$\SmanageB$ &~$\SshipA$ &~$\SshipB$ &~$\ScraneA$ &~$\ScraneB$ &~$\Sallocate$ &~$\Sberth$ &~$\Sdock$ 
		&~$\SoperateA$ &~$\SoperateB$ &~$\Scarrier$ &~$\Sassign$ &~$\Sserve$ &~$\Supdate$ &~$\Sdone$ &~$\ScompleteA$ &~$\ScompleteB$ 
		&~$\SdepartA$ &~$\SdepartB$ \\ \hline\hline
	
	$\Kcrane$ 
		&~$\Kcrane$ &~$\Kcrane$ &~$\Kcrane$ &~$\Kcrane$ &~$\Kcrane$ &~$\Kcrane$ &~$\Kcrane$ &~$\Kplan$ &~$\Kcrane$ &~$\Kcrane$ &~$\Kcrane$ 
		&~$\Kcrane$ &~$\Kcrane$ &~$\Kcrane$ &~$\Kcrane$ &~$\Kcrane$ &~$\Kcrane$ &~$\Kcrane$ &~$\Kcrane$ &~$\Kcrane$ &~$\Kcrane$	\\ \hline	
	
	$\Kplan$ 
		&~$\Kplan$ &~$\Kplan$ &~$\Kplan$ &~$\Kplan$ &~$\Kcrane$ &~$\Kplan$ &~$\Kplan$ &~$\Kplan$ &~$\Kplan$ &~$\Kdock$ &~$\Kplan$ 
		&~$\Kplan$ &~$\Kplan$ &~$\Kplan$ &~$\Kplan$ &~$\Kplan$ &~$\Kplan$ &~$\Kplan$ &~$\Kplan$ &~$\Kplan$ &~$\Kplan$	\\ \hline

	$\Kdock$ 
		&~$\Kdock$ &~$\Kdock$ &~$\Kdock$ &~$\Kdock$ &~$\Kcrane$ &~$\Kdock$ &~$\Kdock$ &~$\Kdock$ &~$\Kdock$ &~$\Kdock$ &~$\Kdock$ 
		&~$\Kdock$ &~$\Kdock$ &~$\Kdock$ &~$\Kdock$ &~$\Kdock$ &~$\Krlse$ &~$\Kdock$ &~$\Kdock$ &~$\Kdock$ &~$\Kdock$	\\ \hline

	$\Krlse$ 
		&~$\Krlse$ &~$\Krlse$ &~$\Krlse$ &~$\Krlse$ &~$\Kcrane$ &~$\Krlse$ &~$\Krlse$ &~$\Krlse$ &~$\Krlse$ &~$\Krlse$ &~$\Krlse$ 
		&~$\Krlse$ &~$\Krlse$ &~$\Krlse$ &~$\Krlse$ &~$\Krlse$ &~$\Krlse$ &~$\Krlse$ &~$\Krlse$ &~$\Krlse$ &~$\Krlse$	\\ \hline

\multicolumn{2}{c}{} \\\hline

	$\outOp$ 
		&~$\Sarrive$ &~$\SmanageA$ &~$\SmanageB$ &~$\SshipA$ &~$\SshipB$ &~$\ScraneA$ &~$\ScraneB$ &~$\Sallocate$ &~$\Sberth$ &~$\Sdock$ 
		&~$\SoperateA$ &~$\SoperateB$ &~$\Scarrier$ &~$\Sassign$ &~$\Sserve$ &~$\Supdate$ &~$\Sdone$ &~$\ScompleteA$ &~$\ScompleteB$ 
		&~$\SdepartA$ &~$\SdepartB$ \\ \hline\hline
		
	$\Kcrane$ 
		&~$\Nstim$ &~$\Nstim$ &~$\Nstim$ &~$\Nstim$ &~$\Nstim$ &~$\Nstim$ &~$\Nstim$ &~$\Sberth$ &~$\Nstim$ &~$\Nstim$ &~$\Nstim$ 
		&~$\Nstim$ &~$\Nstim$ &~$\Nstim$ &~$\Nstim$ &~$\Nstim$ &~$\Nstim$ &~$\Nstim$ &~$\Nstim$ &~$\Nstim$ &~$\Nstim$	\\ \hline	
	
	$\Kplan$ 
		&~$\Nstim$ &~$\Nstim$ &~$\Nstim$ &~$\Nstim$ &~$\ScraneB$ &~$\Nstim$ &~$\Nstim$ &~$\Nstim$ &~$\Nstim$ &~$\SoperateB$ &~$\Nstim$ 
		&~$\Nstim$ &~$\Nstim$ &~$\Nstim$ &~$\Nstim$ &~$\Nstim$ &~$\Nstim$ &~$\Nstim$ &~$\Nstim$ &~$\Nstim$ &~$\Nstim$	\\ \hline

	$\Kdock$ 
		&~$\Nstim$ &~$\Nstim$ &~$\Nstim$ &~$\Nstim$ &~$\ScraneB$ &~$\Nstim$ &~$\Nstim$ &~$\Nstim$ &~$\Nstim$ &~$\Nstim$ &~$\Nstim$ 
		&~$\Nstim$ &~$\Nstim$ &~$\Nstim$ &~$\Nstim$ &~$\Nstim$ &~$\ScompleteB$ &~$\Nstim$ &~$\Nstim$ &~$\Nstim$ &~$\Nstim$	\\ \hline
	
	$\Krlse$ 
		&~$\Nstim$ &~$\Nstim$ &~$\Nstim$ &~$\Nstim$ &~$\ScraneB$ &~$\Nstim$ &~$\Nstim$ &~$\Nstim$ &~$\Nstim$ &~$\Nstim$ &~$\Nstim$ 
		&~$\Nstim$ &~$\Nstim$ &~$\Nstim$ &~$\Nstim$ &~$\Nstim$ &~$\Nstim$ &~$\Nstim$ &~$\Nstim$ &~$\Nstim$ &~$\Nstim$	\\ \hline		

	 \end{tabular}}
\end{table}
\begin{table}[ht!]
	\caption{Stimulus-response specification of the \terminalmanager~$\TM$.}
	\label{tbl:agent_TM}
	\resizebox{1.35\textwidth}{!}{
	\begin{tabular}{|p{1.15cm}||p{1.15cm}|p{1.15cm}|p{1.15cm}|p{1.15cm}|p{1.15cm}|p{1.15cm}|p{1.15cm}|p{1.15cm}|p{1.15cm}|p{1.15cm}
					|p{1.15cm}| p{1.15cm}|p{1.15cm}|p{1.15cm}|p{1.15cm}|p{1.15cm}|p{1.15cm}|p{1.15cm}|p{1.15cm}|p{1.15cm}|p{1.15cm}|} \hline
	$\actOp$ 
		&~$\Sarrive$ &~$\SmanageA$ &~$\SmanageB$ &~$\SshipA$ &~$\SshipB$ &~$\ScraneA$ &~$\ScraneB$ &~$\Sallocate$ &~$\Sberth$ &~$\Sdock$ 
		&~$\SoperateA$ &~$\SoperateB$ &~$\Scarrier$ &~$\Sassign$ &~$\Sserve$ &~$\Supdate$ &~$\Sdone$ &~$\ScompleteA$ &~$\ScompleteB$ 
		&~$\SdepartA$ &~$\SdepartB$ \\ \hline\hline
	
	$\Kallo$ 
		&~$\Kallo$ &~$\Kallo$ &~$\Kallo$ &~$\Kallo$ &~$\Kallo$ &~$\Kallo$ &~$\Kallo$ &~$\Kallo$ &~$\Kallo$ &~$\Kallo$ &~$\Kallo$ 
		&~$\Kallo$ &~$\Kallo$ &~$\Kallo$ &~$\Kallo$ &~$\Kallo$ &~$\Kallo$ &~$\Kfree$ &~$\Kfree$ &~$\Kallo$ &~$\Kallo$	\\ \hline	
	
	$\Kfree$ 
		&~$\Kfree$ &~$\Kfree$ &~$\Kfree$ &~$\Kfree$ &~$\Kfree$ &~$\Kallo$ &~$\Kallo$ &~$\Kfree$ &~$\Kfree$ &~$\Kfree$ &~$\Kfree$ 
		&~$\Kfree$ &~$\Kfree$ &~$\Kfree$ &~$\Kfree$ &~$\Kfree$ &~$\Kfree$ &~$\Kfree$ &~$\Kfree$ &~$\Kfree$ &~$\Kfree$	\\ \hline

\multicolumn{2}{c}{} \\\hline

	$\outOp$ 
		&~$\Sarrive$ &~$\SmanageA$ &~$\SmanageB$ &~$\SshipA$ &~$\SshipB$ &~$\ScraneA$ &~$\ScraneB$ &~$\Sallocate$ &~$\Sberth$ &~$\Sdock$ 
		&~$\SoperateA$ &~$\SoperateB$ &~$\Scarrier$ &~$\Sassign$ &~$\Sserve$ &~$\Supdate$ &~$\Sdone$ &~$\ScompleteA$ &~$\ScompleteB$ 
		&~$\SdepartA$ &~$\SdepartB$ \\ \hline\hline
		
	$\Kallo$ 
		&~$\Nstim$ &~$\Nstim$ &~$\Nstim$ &~$\Nstim$ &~$\Nstim$ &~$\Nstim$ &~$\Nstim$ &~$\Nstim$ &~$\Nstim$ &~$\Nstim$ &~$\Nstim$ 
		&~$\Nstim$ &~$\Nstim$ &~$\Nstim$ &~$\Nstim$ &~$\Nstim$ &~$\Nstim$ &~$\Nstim$ &~$\Nstim$ &~$\Nstim$ &~$\Nstim$	\\ \hline	
	
	$\Kfree$ 
		&~$\Nstim$ &~$\Nstim$ &~$\Nstim$ &~$\Nstim$ &~$\Nstim$ &~$\Sallocate$ &~$\Sallocate$ &~$\Nstim$ &~$\Nstim$ &~$\Nstim$ &~$\Nstim$ 
		&~$\Nstim$ &~$\Nstim$ &~$\Nstim$ &~$\Nstim$ &~$\Nstim$ &~$\Nstim$ &~$\Nstim$ &~$\Nstim$ &~$\Nstim$ &~$\Nstim$	\\ \hline

	 \end{tabular}}
\end{table}
\begin{table}[ht!]
	\caption{Stimulus-response specification of the \crane~$\CM$.}
	\label{tbl:agent_CM}
	\resizebox{1.35\textwidth}{!}{
	\begin{tabular}{|p{1.15cm}||p{1.15cm}|p{1.15cm}|p{1.15cm}|p{1.15cm}|p{1.15cm}|p{1.15cm}|p{1.15cm}|p{1.15cm}|p{1.15cm}|p{1.15cm}
					|p{1.15cm}| p{1.15cm}|p{1.15cm}|p{1.15cm}|p{1.15cm}|p{1.15cm}|p{1.15cm}|p{1.15cm}|p{1.15cm}|p{1.15cm}|p{1.15cm}|} \hline
	$\actOp$ 
		&~$\Sarrive$ &~$\SmanageA$ &~$\SmanageB$ &~$\SshipA$ &~$\SshipB$ &~$\ScraneA$ &~$\ScraneB$ &~$\Sallocate$ &~$\Sberth$ &~$\Sdock$ 
		&~$\SoperateA$ &~$\SoperateB$ &~$\Scarrier$ &~$\Sassign$ &~$\Sserve$ &~$\Supdate$ &~$\Sdone$ &~$\ScompleteA$ &~$\ScompleteB$ 
		&~$\SdepartA$ &~$\SdepartB$ \\ \hline\hline
	
	$\Kread$ 
		&~$\Kread$ &~$\Kread$ &~$\Kread$ &~$\Kread$ &~$\Kread$ &~$\Kread$ &~$\Kread$ &~$\Kread$ &~$\Kread$ &~$\Kread$ &~$\Kread$ 
		&~$\Kread$ &~$\Kread$ &~$\Kseq$ &~$\Kread$ &~$\Kupdate$ &~$\Kread$ &~$\Kread$ &~$\Kread$ &~$\Kread$ &~$\Kread$	\\ \hline	
	
	$\Kcargo$ 
		&~$\Kcargo$ &~$\Kcargo$ &~$\Kcargo$ &~$\Kcargo$ &~$\Kcargo$ &~$\Kcargo$ &~$\Kcargo$ &~$\Kcargo$ &~$\Kcargo$ &~$\Kcargo$ &~$\Kcargo$ 
		&~$\Kcargo$ &~$\Kcargo$ &~$\Kserve$ &~$\Kcargo$ &~$\Koperate$ &~$\Kcargo$ &~$\Kcargo$ &~$\Kcargo$ &~$\Kcargo$ &~$\Kcargo$	\\ \hline
		
	$\Kseq$ 
		&~$\Kseq$ &~$\Kseq$ &~$\Kseq$ &~$\Kseq$ &~$\Kseq$ &~$\Kseq$ &~$\Kseq$ &~$\Kseq$ &~$\Kseq$ &~$\Kseq$ &~$\Kread$ 
		&~$\Kread$ &~$\Kseq$ &~$\Kseq$ &~$\Kseq$ &~$\Kupdate$ &~$\Kseq$ &~$\Kseq$ &~$\Kseq$ &~$\Kseq$ &~$\Kseq$	\\ \hline
		
	$\Kserve$ 
		&~$\Kserve$ &~$\Kserve$ &~$\Kserve$ &~$\Kserve$ &~$\Kserve$ &~$\Kserve$ &~$\Kserve$ &~$\Kserve$ &~$\Kserve$ &~$\Kserve$ &~$\Kcargo$ 
		&~$\Kcargo$ &~$\Kserve$ &~$\Kserve$ &~$\Kserve$ &~$\Koperate$ &~$\Kserve$ &~$\Kserve$ &~$\Kserve$ &~$\Kserve$ &~$\Kserve$	\\ \hline

	$\Kupdate$ 
		&~$\Kupdate$ &~$\Kupdate$ &~$\Kupdate$ &~$\Kupdate$ &~$\Kupdate$ &~$\Kupdate$ &~$\Kupdate$ &~$\Kupdate$ &~$\Kupdate$ &~$\Kupdate$ &~$\Kread$ 
		&~$\Kread$ &~$\Kupdate$ &~$\Kseq$ &~$\Kupdate$ &~$\Kupdate$ &~$\Kupdate$ &~$\Kupdate$ &~$\Kupdate$ &~$\Kupdate$ &~$\Kupdate$	\\ \hline

	$\Koperate$ 
		&~$\Koperate$ &~$\Koperate$ &~$\Koperate$ &~$\Koperate$ &~$\Koperate$ &~$\Koperate$ &~$\Koperate$ &~$\Koperate$ &~$\Koperate$ &~$\Koperate$ &~$\Kcargo$ 
		&~$\Kcargo$ &~$\Koperate$ &~$\Kserve$ &~$\Koperate$ &~$\Koperate$ &~$\Koperate$ &~$\Koperate$ &~$\Koperate$ &~$\Koperate$ &~$\Koperate$	\\ \hline

\multicolumn{2}{c}{} \\\hline

	$\outOp$ 
		&~$\Sarrive$ &~$\SmanageA$ &~$\SmanageB$ &~$\SshipA$ &~$\SshipB$ &~$\ScraneA$ &~$\ScraneB$ &~$\Sallocate$ &~$\Sberth$ &~$\Sdock$ 
		&~$\SoperateA$ &~$\SoperateB$ &~$\Scarrier$ &~$\Sassign$ &~$\Sserve$ &~$\Supdate$ &~$\Sdone$ &~$\ScompleteA$ &~$\ScompleteB$ 
		&~$\SdepartA$ &~$\SdepartB$ \\ \hline\hline
		
	$\Kread$ 
		&~$\Nstim$ &~$\Nstim$ &~$\Nstim$ &~$\Nstim$ &~$\Nstim$ &~$\Nstim$ &~$\Nstim$ &~$\Nstim$ &~$\Nstim$ &~$\Nstim$ &~$\Nstim$ 
		&~$\Nstim$ &~$\Nstim$ &~$\Sassign$ &~$\Nstim$ &~$\Supdate$ &~$\Nstim$ &~$\Nstim$ &~$\Nstim$ &~$\Nstim$ &~$\Nstim$	\\ \hline	
	
	$\Kcargo$ 
		&~$\Nstim$ &~$\Nstim$ &~$\Nstim$ &~$\Nstim$ &~$\Nstim$ &~$\Nstim$ &~$\Nstim$ &~$\Nstim$ &~$\Nstim$ &~$\Nstim$ &~$\Nstim$ 
		&~$\Nstim$ &~$\Nstim$ &~$\Sserve$ &~$\Nstim$ &~$\Sdone$ &~$\Nstim$ &~$\Nstim$ &~$\Nstim$ &~$\Nstim$ &~$\Nstim$	\\ \hline

	$\Kseq$ 
		&~$\Nstim$ &~$\Nstim$ &~$\Nstim$ &~$\Nstim$ &~$\Nstim$ &~$\Nstim$ &~$\Nstim$ &~$\Nstim$ &~$\Nstim$ &~$\Nstim$ &~$\SoperateA$ 
		&~$\SoperateB$ &~$\Nstim$ &~$\Nstim$ &~$\Nstim$ &~$\Supdate$ &~$\Nstim$ &~$\Nstim$ &~$\Nstim$ &~$\Nstim$ &~$\Nstim$	\\ \hline

	$\Kserve$ 
		&~$\Nstim$ &~$\Nstim$ &~$\Nstim$ &~$\Nstim$ &~$\Nstim$ &~$\Nstim$ &~$\Nstim$ &~$\Nstim$ &~$\Nstim$ &~$\Nstim$ &~$\Scarrier$ 
		&~$\Scarrier$ &~$\Nstim$ &~$\Nstim$ &~$\Nstim$ &~$\Sdone$ &~$\Nstim$ &~$\Nstim$ &~$\Nstim$ &~$\Nstim$ &~$\Nstim$	\\ \hline

	$\Kupdate$ 
		&~$\Nstim$ &~$\Nstim$ &~$\Nstim$ &~$\Nstim$ &~$\Nstim$ &~$\Nstim$ &~$\Nstim$ &~$\Nstim$ &~$\Nstim$ &~$\Nstim$ &~$\SoperateA$ 
		&~$\SoperateB$ &~$\Nstim$ &~$\Sassign$ &~$\Nstim$ &~$\Nstim$ &~$\Nstim$ &~$\Nstim$ &~$\Nstim$ &~$\Nstim$ &~$\Nstim$	\\ \hline

	$\Koperate$ 
		&~$\Nstim$ &~$\Nstim$ &~$\Nstim$ &~$\Nstim$ &~$\Nstim$ &~$\Nstim$ &~$\Nstim$ &~$\Nstim$ &~$\Nstim$ &~$\Nstim$ &~$\Scarrier$ 
		&~$\Scarrier$ &~$\Nstim$ &~$\Sserve$ &~$\Nstim$ &~$\Nstim$ &~$\Nstim$ &~$\Nstim$ &~$\Nstim$ &~$\Nstim$ &~$\Nstim$	\\ \hline

	 \end{tabular}}
\end{table}
\begin{table}[ht!]
	\caption{Stimulus-response specification of the \carrier~$\CC$.}
	\label{tbl:agent_CC}
	\resizebox{1.35\textwidth}{!}{
	\begin{tabular}{|p{1.15cm}||p{1.15cm}|p{1.15cm}|p{1.15cm}|p{1.15cm}|p{1.15cm}|p{1.15cm}|p{1.15cm}|p{1.15cm}|p{1.15cm}|p{1.15cm}
					|p{1.15cm}| p{1.15cm}|p{1.15cm}|p{1.15cm}|p{1.15cm}|p{1.15cm}|p{1.15cm}|p{1.15cm}|p{1.15cm}|p{1.15cm}|p{1.15cm}|} \hline
	$\actOp$ 
		&~$\Sarrive$ &~$\SmanageA$ &~$\SmanageB$ &~$\SshipA$ &~$\SshipB$ &~$\ScraneA$ &~$\ScraneB$ &~$\Sallocate$ &~$\Sberth$ &~$\Sdock$ 
		&~$\SoperateA$ &~$\SoperateB$ &~$\Scarrier$ &~$\Sassign$ &~$\Sserve$ &~$\Supdate$ &~$\Sdone$ &~$\ScompleteA$ &~$\ScompleteB$ 
		&~$\SdepartA$ &~$\SdepartB$ \\ \hline\hline
	
	$\Kavail$ 
		&~$\Kavail$ &~$\Kavail$ &~$\Kavail$ &~$\Kavail$ &~$\Kavail$ &~$\Kavail$ &~$\Kavail$ &~$\Kavail$ &~$\Kavail$ &~$\Kavail$ &~$\Kavail$ 
		&~$\Kavail$ &~$\Kavail$ &~$\Kavail$ &~$\Knear$ &~$\Kavail$ &~$\Kavail$ &~$\Kavail$ &~$\Kavail$ &~$\Kavail$ &~$\Kavail$	\\ \hline	
	
	$\Kassign$ 
		&~$\Kassign$ &~$\Kassign$ &~$\Kassign$ &~$\Kassign$ &~$\Kassign$ &~$\Kassign$ &~$\Kassign$ &~$\Kassign$ &~$\Kassign$ &~$\Kassign$ &~$\Kassign$ 
		&~$\Kassign$ &~$\Kassign$ &~$\Kassign$ &~$\Kmove$ &~$\Kassign$ &~$\Kassign$ &~$\Kassign$ &~$\Kassign$ &~$\Kassign$ &~$\Kassign$	\\ \hline

	$\Knear$ 
		&~$\Knear$ &~$\Knear$ &~$\Knear$ &~$\Knear$ &~$\Knear$ &~$\Knear$ &~$\Knear$ &~$\Knear$ &~$\Knear$ &~$\Knear$ &~$\Knear$ 
		&~$\Knear$ &~$\Kavail$ &~$\Knear$ &~$\Knear$ &~$\Knear$ &~$\Knear$ &~$\Knear$ &~$\Knear$ &~$\Knear$ &~$\Knear$	\\ \hline

	$\Kmove$ 
		&~$\Kmove$ &~$\Kmove$ &~$\Kmove$ &~$\Kmove$ &~$\Kmove$ &~$\Kmove$ &~$\Kmove$ &~$\Kmove$ &~$\Kmove$ &~$\Kmove$ &~$\Kmove$ 
		&~$\Kmove$ &~$\Kassign$ &~$\Kmove$ &~$\Kmove$ &~$\Kmove$ &~$\Kmove$ &~$\Kmove$ &~$\Kmove$ &~$\Kmove$ &~$\Kmove$	\\ \hline

\multicolumn{2}{c}{} \\\hline

	$\outOp$ 
		&~$\Sarrive$ &~$\SmanageA$ &~$\SmanageB$ &~$\SshipA$ &~$\SshipB$ &~$\ScraneA$ &~$\ScraneB$ &~$\Sallocate$ &~$\Sberth$ &~$\Sdock$ 
		&~$\SoperateA$ &~$\SoperateB$ &~$\Scarrier$ &~$\Sassign$ &~$\Sserve$ &~$\Supdate$ &~$\Sdone$ &~$\ScompleteA$ &~$\ScompleteB$ 
		&~$\SdepartA$ &~$\SdepartB$ \\ \hline\hline
		
	$\Kavail$ 
		&~$\Nstim$ &~$\Nstim$ &~$\Nstim$ &~$\Nstim$ &~$\Nstim$ &~$\Nstim$ &~$\Nstim$ &~$\Nstim$ &~$\Nstim$ &~$\Nstim$ &~$\Nstim$ 
		&~$\Nstim$ &~$\Nstim$ &~$\Nstim$ &~$\Sserve$ &~$\Nstim$ &~$\Nstim$ &~$\Nstim$ &~$\Nstim$ &~$\Nstim$ &~$\Nstim$	\\ \hline	
	
	$\Kassign$ 
		&~$\Nstim$ &~$\Nstim$ &~$\Nstim$ &~$\Nstim$ &~$\Nstim$ &~$\Nstim$ &~$\Nstim$ &~$\Nstim$ &~$\Nstim$ &~$\Nstim$ &~$\Nstim$ 
		&~$\Nstim$ &~$\Nstim$ &~$\Nstim$ &~$\Supdate$ &~$\Nstim$ &~$\Nstim$ &~$\Nstim$ &~$\Nstim$ &~$\Nstim$ &~$\Nstim$	\\ \hline

	$\Knear$ 
		&~$\Nstim$ &~$\Nstim$ &~$\Nstim$ &~$\Nstim$ &~$\Nstim$ &~$\Nstim$ &~$\Nstim$ &~$\Nstim$ &~$\Nstim$ &~$\Nstim$ &~$\Nstim$ 
		&~$\Nstim$ &~$\Scarrier$ &~$\Nstim$ &~$\Nstim$ &~$\Nstim$ &~$\Nstim$ &~$\Nstim$ &~$\Nstim$ &~$\Nstim$ &~$\Nstim$	\\ \hline

	$\Kmove$ 
		&~$\Nstim$ &~$\Nstim$ &~$\Nstim$ &~$\Nstim$ &~$\Nstim$ &~$\Nstim$ &~$\Nstim$ &~$\Nstim$ &~$\Nstim$ &~$\Nstim$ &~$\Nstim$ 
		&~$\Nstim$ &~$\Sassign$ &~$\Nstim$ &~$\Nstim$ &~$\Nstim$ &~$\Nstim$ &~$\Nstim$ &~$\Nstim$ &~$\Nstim$ &~$\Nstim$	\\ \hline

	 \end{tabular}}
\end{table}
\end{landscape}

\subsection{\levelTHREE{s}}
\label{sub:concrete_behaviour_specifications}

The \levelThree{s} for the system agents are given in Figs.~\ref{fig:agent_PC}--\ref{fig:agent_CC}. 
Note that in the \levelThree{s}, we use functions as a simplification of the specification in places where we are not concerned with how the system performs the operation. In these cases, we assume that the function arguments are passed by value, meaning that they can only be referenced, and not defined, in the function. For example, in the \levelThree of the \carrier~$\CC$, there is a function call denoted by~$\mathsf{ASSIGN}(\var{carriers},\var{containers})$. In this case, the concrete behaviour references the variables~$\var{carriers}$ and~$\var{containers}$ when determining the straddle carrier assignment. Additionally, note that program variables appearing in all capitals (\eg \var{SHIP\_MANIFEST}, \var{SHIP\_LENGTH}, \var{CRANE\_EFF}, \etc) denote defined constants.

\begin{figure}[ht!]
	\centering
	\hspace{-2.5em}
	\scalebox{0.85}{\parbox{\linewidth}{
	\begin{eqnarray*}
		\KmanA		&\deq& 	\varL{m1} := \true; \varL{i} := 1									\\
		\KmanB		&\deq& 	\varL{m2} := \true; \varL{i} := 2									\\
		\Kinit		&\deq& 	\guardif{(i = 1)}{}											
										\varL{manifest[1]} := \varL{SHIP\_MANIFEST}; 			
										\varL{length[1]} := \varL{SHIP\_LENGTH}; 				\\
					&	 &	\dent{1} 	\varL{numBays[1]} := \varL{SHIP\_BAYS};
										\varL{numContainers[1]} := \varL{SHIP\_CONTAINERS}; 	\\
	 				&	 &	\dent{1}	\varL{arriveT[1]} := \varL{ARRIVE\_TIME}; 			
										\varL{departT[1]} := \varL{DEPART\_TIME}; 		
										\varL{waitT[1]} := \varL{WAIT\_TIME} 					\\
					&	 &	\guardb{(i = 2)}{}												
										\varL{manifest[2]} := \varL{SHIP\_MANIFEST}; 			
										\varL{length[2]} := \varL{SHIP\_LENGTH}; 				\\
					&	 &	\dent{1} 	\varL{numBays[2]} := \varL{SHIP\_BAYS};
										\varL{numContainers[2]} := \varL{SHIP\_CONTAINERS}; 	\\
	 				&	 &	\dent{1}	\varL{arriveT[2]} := \varL{ARRIVE\_TIME}; 			
										\varL{departT[2]} := \varL{DEPART\_TIME}; 			
										\varL{waitT[2]} := \varL{WAIT\_TIME} 					\\
					&	 &	\guarde															\\
		\KclearA	&\deq& 	\varL{m1} := \false; \varL{i} := 1								\\
		\KclearB	&\deq& 	\varL{m2} := \false; \varL{i} := 2								\\
		\Kdepart	&\deq& 	\guardif{(i = 1)}{}												
										\varL{manifest[1]} := \nulls; 						
										\varL{length[1]} := 0; 								
										\varL{numBays[1]} := 0; 								\\
					&	 &	\dent{1} 	\varL{numContainers[1]} := 0; 						
										\varL{arriveT[1]} := 0; 							
										\varL{departT[1]} := 0; 							
										\varL{waitT[1]} := 0 								\\
					&	 &	\guardb{(i = 2)}{}												
										\varL{manifest[2]} := \nulls; 						
										\varL{length[2]} := 0; 								
										\varL{numBays[2]} := 0; 								\\
					&	 &	\dent{1} 	\varL{numContainers[2]} := 0; 						
										\varL{arriveT[2]} := 0; 							
										\varL{departT[2]} := 0;
										\varL{waitT[2]} := 0 								\\
					&	 &	\guarde
	\end{eqnarray*}}}
	
	\caption{Concrete behaviour specification of the \portcaptain~$\PC$ behaviours.}
	\label{fig:agent_PC}
\end{figure}
\begin{figure}[ht!]
	\centering
	\scalebox{0.85}{\parbox{\linewidth}{
	\begin{eqnarray*}
		\KtimeS		&\deq& 	\varL{serviceT[i]} := \varL{departT[i]} - \varL{arriveT[i]} - \varL{waitT[i]}	\\
		\Kpos		&\deq& 	\varL{dockPos[i]} := \varL{berthPos[i]}										\\
		\Kleave	&\deq& 	\varL{dockPos[i]} := \nulls; \varL{serviceT[i]} := 0							
	\end{eqnarray*}}}
	
	\caption{Concrete behaviour specification of the \shipmanager behaviours where~$i = 1$ for~$\SM{1}$ and~$i = 2$ for~$\SM{2}$.}
	\label{fig:agent_SM}
\end{figure}
\begin{figure}[ht!]
	\centering
	\scalebox{0.85}{\parbox{\linewidth}{
	\begin{eqnarray*}
		\Kcrane	&\deq& 	\varL{numCranes[i]} := \varL{numContainers[i]} / (\varL{CRANE\_EFF} * \varL{serviceT[i]})	\\
		\Kplan		&\deq& 	\varL{berthPos[i]} := \varL{berth[i]}; 													\\
					&	 &	\varL{bayPlan[i]} :=\mathsf{PLAN}(\varL{berthPos[i]},\varL{alloCranes[i]},\varL{manifest[i]},\varL{numBays[i]})\\
		\Kdock	&\deq& 	\varL{docked[i]} := \true 																\\
		\Krlse		&\deq& 	\varL{docked[i]} := \false;															  
					\varL{berthPos[i]} := \nulls;															
					\varL{bayPlan[i]} := \nulls;																
					\varL{numCranes[i]} := 0																
	\end{eqnarray*}}}
	
	\caption{Concrete behaviour specification of the \stevedore behaviours where~$i = 1$ for~$\SV{1}$ and~$i = 2$ for~$\SV{2}$.}
	\label{fig:agent_SV}
\end{figure}
\begin{figure}[ht!]
	\centering
	\scalebox{0.85}{\parbox{\linewidth}{
	\begin{eqnarray*}
		\Kallo	&\deq& 	\receive y;																\\
				&	 &	\guardif{(y \ge \ScraneA)}{}										
									\varL{berth[1]} := \mathsf{POSITION}(\varL{numCranes[1]});	\\
				&	 &	\dent{1} 	\varL{alloCranes[1]} := \mathsf{ALLOCATE}(\varL{berth[1]})	\\
				&	 &	\guardb{(y \ge \ScraneB)}{}												
									\varL{berth[2]} := \mathsf{POSITION}(\varL{numCranes[2]});	\\
				&	 &	\dent{1} \varL{alloCranes[2]} := \mathsf{ALLOCATE}(\varL{berth[2]})		\\
				&	 &	\guarde																	\\
		\Kfree	&\deq& 	\receive y;																\\
				&	 &	\guardif{(y \ge \ScompleteA)}{} 										
									\varL{berth[1]} := \nulls; 									
									\varL{alloCranes[1]} := \nulls 								\\
				&	 &	\guardb{(y \ge \ScompleteB)}{} 										
									\varL{berth[2]} := \nulls; 									
									\varL{alloCranes[2]} := \nulls 								\\
				&	 &	\guarde																		
	\end{eqnarray*}}}
	
	\caption{Concrete behaviour specification of the \terminalmanager~$\TM$ behaviours.}
	\label{fig:agent_TM}
\end{figure}
\begin{figure}[ht!]
	\centering
	\scalebox{0.85}{\parbox{\linewidth}{
	\begin{eqnarray*}
		\Kread		&\deq& 	\receive y;														\\
					&	 &	\guardif{(y \ge \SoperateA)}{\varL{plan} := \varL{bayPlan[1]}}	\\
					&	 &	\guardb{(y \ge\SoperateB)}{\varL{plan} := \varL{bayPlan[2]}}		\\
					&	 &	\guarde															\\
		\Kcargo		&\deq& 	\varL{containers} := \mathsf{CONTAINERS}(\varL{plan})				\\
		\Kseq		&\deq& 	\varL{sequence} := \mathsf{SEQUENCE}(\varL{carrierAssign})		\\
		\Kserve		&\deq& 	\varL{position} := \mathsf{SERVICE}(\varL{sequence})				\\
		\Kupdate	&\deq& 	\varL{plan} := \mathsf{UPDATE}(\varL{carrierState})				\\
		\Koperate	&\deq& 	\varL{operation} := \mathsf{OPERATE}(\varL{plan})	
	\end{eqnarray*}}}
	
	\caption{Concrete behaviour specification of the \crane~$\CM$ behaviours.}
	\label{fig:agent_CM}
\end{figure}
\begin{figure}[ht!]
	\centering
	\scalebox{0.85}{\parbox{\linewidth}{
	\begin{eqnarray*}
		\Kavail		&\deq& 	\varL{carriers} := \mathsf{AVAIL}(\varL{carrierState})									\\
		\Kassign	&\deq& 	\varL{carrierAssign} := \mathsf{ASSIGN}(\varL{carriers},\varL{containers})					\\
		\Knear		&\deq& 	\varL{nearest} := \mathsf{NEAREST}(\varL{carriers},\varL{position})						\\
		\Kmove		&\deq& 	\varL{carrierState} := \mathsf{MOVE}(\varL{carrierState},\varL{nearest},\varL{position})
	\end{eqnarray*}}}
	
	\caption{Concrete behaviour specification of the \carrier~$\CC$ behaviours.}
	\label{fig:agent_CC}
\end{figure}


\pagebreak

\end{document}